\documentclass[journal,onecolumn,draftclsnofoot]{IEEEtran}
\IEEEoverridecommandlockouts
\usepackage{t1enc}
\usepackage[utf8x]{inputenc}
\usepackage[english]{babel}
\usepackage{pdfrack}
\usepackage{psfrag}
\usepackage{dsfont}
\usepackage{amsmath}
\usepackage{amssymb}
\usepackage{stmaryrd}
\usepackage[amsthm]{ntheorem-hyper}
\usepackage{flushend}
\usepackage{graphicx}
\usepackage{subfloat}
\usepackage{subfig}
\usepackage{caption}
\usepackage{float}
\usepackage{enumerate}

\def\iX{{\cal X}}

\def\iP{{\cal P}}

\newcommand{\sumfrac}[2]{\genfrac{}{}{0pt}{}{#1}{#2}}

\newcommand{\Prob}{\textnormal{Pr}}

\theoremstyle{plain}
\newtheorem{Thm}{Theorem}

\newtheorem{Lem}{Lemma}
\newtheorem{Cor}{Corollary}
\theoremstyle{definition}
\newtheorem{Def}{Definition}

\theoremstyle{remark}
\newtheorem{Rem}{Remark}

\newcommand{\vx}{\mathbf x}
\newcommand{\vy}{\mathbf y}

\DeclareMathOperator{\I}{I}
\DeclareMathOperator{\HH}{H}
\DeclareMathOperator{\DD}{D}
\DeclareMathOperator{\D}{D}

\DeclareMathOperator{\J}{J}
\title{Universal Random Access Error Exponents for Codebooks of Different Word-Lengths}
\author{L\'or\'ant Farkas and
        Tamás Kói
\thanks{This paper has been presented in part at the recent result poster session of ISIT 2016, Barcelona. L\'or\'ant Farkas is with the Department of Analysis, Budapest University of Technology and Economics,
e-mail: lfarkas@math.bme.hu. Tamás Kói is with the Department of Stochastics, Budapest University of Technology and Economics,
e-mail: koitomi@math.bme.hu. The work of the authors was supported by the Hungarian National Research Development and Innovation Office Grant K105840.}}

\begin{document}

\maketitle
\begin{abstract}
Csisz\'ar's channel coding theorem for multiple codebooks is generalized allowing the codeword lenghts differ across codebooks. Also in this case, for each codebook an error exponent can be achieved that equals the random coding exponent for this codebook alone, in addition, erasure detection failure probability tends to $0$. This is proved even for sender and receiver not knowing the channel. As a corollary, a substantial improvement is obtained when the sender knows the channel. 
\end{abstract}


\section{Introduction}
The discrete memoryless channel (DMC) coding theorem of Csisz\'ar \cite{Csiszar} analyzes the performance of a codebook library of several constant composition codebooks consisting of codewords of the same length. The rate and the type of the codewords may be different for each codebook. The number of codebooks is subexponential in the codeword length. It is shown that simultaneously for each codebook the same error exponent can be achieved as the random coding exponent of this codebook alone. In other words, for transmitting messages that may be of different kinds, with specified rates: with the sender using different codebooks for different kinds of messages, the same reliability can be guaranteed for each message kind as if it were known that only messages of this kinds occur, with the given rates. Note that this theorem is used in \cite{Csiszar} to the engineeringly different problem of joint source-channel coding. As noted in \cite{Csiszar} the result is also relevant in unequal protection of messages: for better protection, important messages may be encoded via "more reliable" (smaller) codebooks, see for example Borade, Nakiboglu and Zheng \cite{Borade}, Weinberger and Merhav \cite{Merhav} and Shkel, Tan and Draper \cite{Yanina} for more recent results.

Luo and Ephremides in \cite{Ephremides} analyze a similar model in the context of random access communication for multiple access channel (MAC) which brings classical information theory closer to packet based random access communications models. Not using their concepts of standard communication rate and generalized random coding, the model of \cite{Ephremides} can be summarized as follows. Each user employs a random codebook partitioned into classes corresponding to different rate options. If the vector of the senders' actual rate choices belongs to a preselected operation rate region, the decoder should reliably
decode the messages sent, otherwise it should report collision. Wang and Luo in \cite{Jie-Luo} derive Gallager type error exponents for this model. In a slightly modified model Farkas and K\'oi in \cite{isit2013} give error exponents employing a mutual information based universal decoder, with application to joint source-channel coding for MAC.

This paper generalizes the mentioned result of \cite{Csiszar} in a different direction, not addressing MACs. As in \cite{Csiszar}, the sender is assumed to have a codebook library of several codebooks, each consisting of codewords of the same length and type. Before each message transmission, the sender chooses the codebook he will use, the receiver is unaware of this choice. As a new feature compared to \cite{Csiszar}, here not only the rate and type but also the codeword length may vary across codebooks, thus a model in between fixed and variable length coding is addressed. This model appears natural, e.g., for communication situations where a channel is used alternatingly for transmitting messages of different kinds such as audio, data, video etc. We believe that this paper, though of theoretical nature providing asymptotic achievability results, may contribute to a better understanding of such communication situations.

For channels with positive zero-error capacity, the above model does not provide mathematical challenges. Indeed, in that case (as noted also in \cite{Csiszar}) prior to each message transmission the sender can communicate his codebook choice over the channel without error, using codewords of length $o(n)$. This reduces the introduced model to the standard case of a single codebook.

In the more common case of zero error capacity equal to $0$, no such simple strategy is available, and the fact that codewords of different length are used causes a certain asynchronism at the receiver, who should also estimate the boundaries of the codewords and avoid error propagation. To meet these challenges we introduce a mutual information based two-stage decoder.

It is not obvious what to mean by decoding error in our model. By the definition we adopt, the $j$'th message is correctly decoded if the decoder correctly assigns this message to the time slot where the corresponding codeword is sent, including correct identification of the codeword boundaries. The receiver is not required to learn that this message has been sent as the $j$'th one (taking care of the possibility that at previous instances erroneously less or more messages have been decoded than actually sent).

Our main result extends the result in \cite{Csiszar} to the above scenario, showing that simultaneously for each codebook choice the same error exponent can be achieved as the random coding error exponent for the chosen codebook alone. This is proved under the technical assumption that all codeword length ratios are between $D$ and $\frac{1}{D}$ for some $D \in (0,1]$ and the number of codebooks is subexponential in length-bound $n$. Recall that even in the standard case of a single codebook, a positive error exponent is achievable only for rate less than the mutual information over the channel with input distribution equal to the type of the codewords, and under this condition the random coding exponent is positive. It is desirable that when this condition fails, the decoder can report that reliable decoding is not possible. This feature is present in \cite{Ephremides} and \cite{isit2013} (but not in \cite{Csiszar}). In \cite{Ephremides} and \cite{isit2013}, addressing MACs, the term collision detection is used, in our one-sender context we will use the term erasure detection. As part of main result, our universal decoder is shown suitable also for erasure detection: When the chosen codebook has random coding exponent $0$, an erasure is reported with probability approaching $1$, though here we do not obtain exponential speed of convergence (for more on this see Remark \ref{etan}). This has been achieved with a completely universal construction: Neither the design of the codebook library nor the decoder depends on the channel.

A corollary of the main theorem improves the result when the sender knows the channel while maintaining the universality of the decoder. The improvement leads to exponent also for erasure declaration failure probability, and shows that for each message kind the maximum of the random coding error exponent over the possible input distributions is achievable. Even the special case of this corollary for transmitting messages of a single kind is of interest, yielding a universal coding  result for this classical problem that, to our knowledge, does not appear in the literature, see Remark \ref{singlecodebook}.

The proofs rely on the subtype technique of Farkas and K\'oi \cite{Hawaii} and \cite{Hongkong}. The hardest kind of error to deal with has been that of detecting the right codeword in a wrong position, partially overlapping with the correct one. This obstacle has been overcome employing a new concept of $\gamma$-independent sequences, and also second order types.

We are aware of only one prior work extending results in \cite{Csiszar} in a direction like here, by Balakirsky \cite{Balakirsky} on joint source-channel coding error exponent for variable length codes. Channel coding with multiple codebooks is not explicitly mentioned in \cite{Balakirsky} but some ideas in our paper are similar to those there, due to the close mathematical relationship of these problems.

Note that the topic of the paper is also connected (see the Discussion for details) to the area of strong asynchronism, see Tchamkerten, Chandar and Wornell \cite{Tchamkarten} and Polyanskiy \cite{Polyanskiy}, and even more to Yıldırım, Martinez and Fàbregas \cite{HongkongML} concerning error exponents.

\section{Notation} \label{notation}
The notation follows \cite{Csiszar}, \cite{Nazari} and \cite{Hawaii} whenever possible. All alphabets are finite and $\log$ denotes logarithm to the base $2$. The set $\{1,2,\dots,M\}$ is denoted by $[M]$. The notation $subexp(n)$ denotes a quantity growing subexponentially as $n \rightarrow \infty$  (i.e. $\frac{1}{n}\log(subexp(n))\rightarrow 0$), that could be given explicitly. For some subexpontial sequences individual notations are used and the parameters on which these sequences depend will be indicated in parantheses.

Random variables $X$, $Y$, etc., with alphabets $\mathcal{X}$, $\mathcal{Y}$, etc., will be assigned several different (joint) distributions. These will be denoted by $P^{X}$, $P^{XY}$, etc. or $V^{X}$, $V^{XY}$, etc. The first notation will typically refer to a distinguished (joint) distribution, the second one refers to distributions introduced for technical purposes such as representing joint types. The family of all distributions on $\mathcal{X} \times \mathcal{Y}$, say, is denoted by $\mathcal{P} (\mathcal{X} \times \mathcal{Y})$. If a multivariate distribution, say $V^{\hat{X}XY} \in \mathcal{P} (\mathcal{X} \times \mathcal{X} \times \mathcal{Y})$ is given then $V^{X}$, $V^{\hat{X}X}$, $V^{XY}$, $V^{Y|X}$  etc. will denote the associated marginal or conditional distributions.

The type of an $n$-length sequence $\vx=x_1 x_2 \dots x_n \in \mathcal{X}^n$ is the distribution $P_{\vx} \in \mathcal{P} (\mathcal{X})$ where $P_{\vx}(x)$ is the relative frequency of the symbol $x$ in $\vx$. The joint type of two or more $n$-length sequences is defined similarly and, for $(\vx,\vy) \in \mathcal{X}^n \times \mathcal{Y}^n$, say, it is denoted by $P_{(\vx,\vy)}$. The family of all possible types of sequences $\vx \in \mathcal{X}^n$ is denoted by $\mathcal{P}^n (\mathcal{X})$, and for $P \in \mathcal{P}^n (\mathcal{X})$  the set of all $\vx \in \mathcal{X}^n$ of type $P_{\vx}=P$ is denoted by $T^n_{P}$.

Denote $\HH_V(X)$, $\HH_V(Y|X)$, $\I_V(\hat{X}X \wedge Y)$ etc. the entropy, conditional entropy and mutual information etc. when the random variables $X$, $\hat{X}$, $Y$ have joint distribution $V=V^{\hat{X}XY}$. Furthermore, the empirical mutual information $\I(\vx \wedge \vy)$ of two sequences $\vx$ and $\vy$ (of equal length) is defined as $\I_V(X \wedge Y)$ with $V^{XY}=P_{(\vx,\vy)}$.

Given a DMC $W: \mathcal{X} \rightarrow \mathcal{Y}$ and $P \in \mathcal{P}(\mathcal{X})$ let $I(P,W)$ be equal to $\I_{V}(X\wedge Y)$ where $V^{X}=P$ and $V^{Y|X}=W$. The maximum of $I(P,W)$ over all $P \in \mathcal{P}(\mathcal{X})$ is the capacity of the DMC $W$.

The following elementary facts will be used (see, e.g., \cite{Csiszar2}):

\vspace{-5mm}

\begin{align}
&|\iP^n(\iX)|\leq (n+1)^{|\iX|},  \label{basicfact1}\\
&\frac{2^{n\HH(P)}}{(n+1)^{|\iX|}} \le |T^n_{P}|\le 2^{n\HH(P)} \textnormal{ if }\,  P \in \iP^n(\iX), \label{basicfact2}\\
&W^n(\vy|\vx)=2^{-n\left(\DD(V^{Y|X}\|W|P_\vx)+\HH_{V}(Y|X)\right)} \textnormal{ where $V^{XY}=P_{(\vx,\vy)}$.} \label{basicfact4}
\end{align}

The concatenation of an $n_1$-type $V_1\in \iP^{n_1}(\iX)$ and an $n_2$-type $V_2\in \iP^{n_2}(\iX)$ is the $(n_1+n_2)$-type  $V_1\oplus V_2\in \iP^{n_1+n_2}(\iX)$ with
 \begin{align}
 \left( V_1\oplus V_2 \right)(x)=\frac{n_1}{n_1+n_2}V_1(x)+\frac{n_2}{n_1+n_2}V_2 (x).
 \end{align}
The concatenation of joint types, say, $V_1 \in \mathcal{P}^{n_1}(\mathcal{X} \times \mathcal{Y})$ and $V_2 \in \mathcal{P}^{n_2}(\mathcal{X} \times \mathcal{Y})$ is defined similarly. If $V_1$, $V_2$, \dots, $V_k$ are $n_1$, $n_2$, \dots, $n_k$-types, respectively, let
 \begin{equation}
 \J(V_1,V_2, \dots, V_k)=\HH(V_1 \oplus \dots \oplus V_k)-\sum_{i=1}^{k}\frac{n_i}{n_1+\dots+n_k}\HH(V_i). \label{F-def1}
 \end{equation}
The nonegative quantity in (\ref{F-def1}) is a Jensen-Shannon divergence if $k=2$, and a generalized Jensen-Shannon divergence otherwise, in the sense of \cite{BURBEA} and \cite{Lin}.

The second order type of a sequence $\mathbf{x}=x_1 \dots x_n \in \mathcal{X}^n$ is $P_{\mathbf{x}}^2 \in \mathcal{P}^{n-1}(\mathcal{X}\times \mathcal{X})$ defined by
\begin{equation}
P_{\mathbf{x}}^2 (a,b) = \frac{1}{n-1}|i: x_i = a, x_{i+1}=b|.
\end{equation}
In other words, $P_{\mathbf{x}}^2$ is the joint type of $\mathbf{x}'=x_1 \dots x_{n-1}$ and $\mathbf{x}''=x_2 \dots x_{n}$. Let $T^{n,2}_{V,a}$ denote the second order type class $\{\mathbf{x}: \mathbf{x}\in \mathcal{X}^n, P_{\mathbf{x}}^2=V,x_1=a \}$. We cite from \cite{Csiszar3} that
\begin{equation}
|T^{n,2}_{V,a}|\le 2^{n\HH_{V}(\hat{X}|X)}. \label{markovrabecsles}
\end{equation}

\begin{figure}[H]
\begin{center}
\includegraphics[scale=1.5]{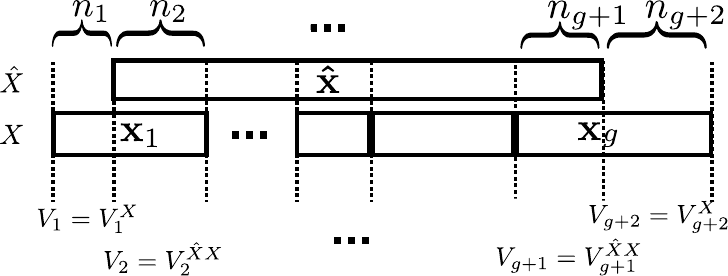}
\caption{Illustration for understanding some notations}
\label{notionalakul}
\end{center}
\end{figure}

The next combinatorial construction will be substantially used in our proofs. Let a $(g+1)$-length sequence of positive integers $\mathbf{L}=(\hat{l},l^1,l^2,\dots, l^g)$, a non-negative integer $q$ and a collection of sequences $(\mathbf{\hat{x}},\mathbf{x}_1, \dots, \mathbf{x}_g)$ with $\mathbf{\hat{x}} \in \mathcal{X}^{\hat{l}}$, $\mathbf{x}_1 \in \mathcal{X}^{l^1}$, $\mathbf{x}_2 \in \mathcal{X}^{l^2}$, $\dots$, $\mathbf{x}_g \in \mathcal{X}^{l^g}$ be given. The sequences $\mathbf{\hat{x}},\mathbf{x}_1, \dots, \mathbf{x}_g$  are arranged in a two-row array as in Figure \ref{notionalakul}, i.e., $\mathbf{\hat{x}}$ is placed in the first row and $\mathbf{x}_1, \dots, \mathbf{x}_g$ are placed consecutively in the second row so that the second row ends by $q$ symbols after the first one; either row may start before the other one, depending on $\mathbf{L}$ and $q$. This configuration is referred to as $(\mathbf{L},q)$-array in the sequel. It will be always assumed that $\mathbf{\hat{x}}$ has a nonempty overlap with both $\mathbf{x}_1$ and $\mathbf{x}_g$, equivalently that
\begin{equation}
q < l^g, \text{  and } \sum_{i=2}^g l^i -q < \hat{l}. \label{szimmetrikusfeltetel}
\end{equation}
Note that the second inequality in (\ref{szimmetrikusfeltetel}) trivially holds if $g=1$.

An $(\mathbf{L},q)$-array is divided into subblocks according to the starting and ending positions of the sequences $\mathbf{\hat{x}},\mathbf{x}_1, \dots, \mathbf{x}_g$  (see Fig. \ref{notionalakul}). For technical reasons, we assume artificially that in the degenerate case of $q=0$ there is a $0$-length last block and  in case of $\hat{l}=\sum_{i=1}^{g} l^i-q$ there is a $0$-length first block. Then the number of the subblocks is always equal to $g+2$. Their lengths, determined by $\mathbf{L}$ and $q$, will be denoted by $n_1, \dots, n_{g+2}$. Note that $q=n_{g+2}$. For $2 \le i \le g+1$, the $i$'th subblock consists of parts both in the first and the second row, let $V_i \in \mathcal{P}^{n_i}(\mathcal{X} \times \mathcal{X})$ denote their joint type. The first and last subblocks are contained in one row, their types are $V_1 \in \mathcal{P}^{n_1}(\mathcal{X})$ and $V_{g+2} \in \mathcal{P}^{n_1}(\mathcal{X})$. The subblock types $V_i$ will be often represented via dummy random variables, with $\hat{X}$ referring to the first and $X$ to the second row. When $V_i=V_i^{\hat{X}X}$, $V_i^{\hat{X}}$ and $V_i^X$ are the types of the parts of the $i$'th subblock in the first resp. second row. In the degenerate case when $n_1=0$ or $n_{g+2}=0$, let $V_1$ resp. $V_{g+2}$ be a dummy symbol regarded as the type of the empty sequence.

Given $\mathbf{L}=(\hat{l},l^1,l^2,\dots, l^g)$ and $q$ satisfying (\ref{szimmetrikusfeltetel}), each sequence $\mathbf{V}=(V_1,\dots,V_{g+2})$ of types $V_1 \in \mathcal{P}^{n_1} (\mathcal{X})$, $V_i \in \mathcal{P}^{n_1} (\mathcal{X} \times \mathcal{X})$, $i=2,\dots,g+1$, and $V_{g+2} \in \mathcal{P}^{n_{g+2}} (\mathcal{X})$, where $n_1, \dots, n_{g+2}$ are the subblock lengths determined by $(\mathbf{L},q)$, will be called a subtype sequence compatible with $(\mathbf{L},q)$. For $\mathbf{V}$ compatible with $(\mathbf{L},q)$, and sequences $\mathbf{\hat{x}} \in \mathcal{X}^{\hat{l}}$, $\mathbf{x}_i \in \mathcal{X}^{l^i}$, $i=2,\dots,g$ let $\mathds 1^{\mathbf{L},q}_{\mathbf{V}}(\mathbf{\hat{x}};\mathbf{x}_1, \dots, \mathbf{x}_g)$ denote the indicator function equal to $1$ if $\mathbf{\hat{x}},\mathbf{x}_1, \dots, \mathbf{x}_g$ arranged in $(\mathbf{L},q)$ array has subtype sequence $\mathbf{V}$, and otherwise $0$. The set of collections of sequences $(\mathbf{\hat{x}},\mathbf{x}_1, \dots, \mathbf{x}_g)$ with $\mathds 1^{\mathbf{L},q}_{\mathbf{V}}(\mathbf{\hat{x}};\mathbf{x}_1, \dots, \mathbf{x}_g)=1$, i.e., for which the corresponding $(\mathbf{L},q)$-array has subtype sequence $\mathbf{V}$, will be denoted by $\mathcal{T}^{\mathbf{L},q}_{\mathbf{V}}$. In the sequel the following generalization of $\mathcal{T}^{\mathbf{L},q}_{\mathbf{V}}$ is also needed. Let $\mathcal{I}$ be a set of prescribed equalities of form $\mathbf{x}_i=\mathbf{x}_j$ with $i,j \in [g]$, or $\mathbf{\hat{x}}=\mathbf{x}_1$ or $\mathbf{\hat{x}}=\mathbf{x}_g$ (the possibility of $\mathbf{\hat{x}}=\mathbf{x}_i$ for $1<i<g$ is excluded since $\mathbf{\hat{x}}$ has nonempty overlap with both $\mathbf{x}_1$ and $\mathbf{x}_g$). The set of those collections $(\mathbf{\hat{x}},\mathbf{x}_1, \dots, \mathbf{x}_g) \in \mathcal{T}^{\mathbf{L},q}_{\mathbf{V}}$ for which the equalities in $\mathcal{I}$ hold will be denoted by $\mathcal{T}^{\mathbf{L},q}_{\mathbf{V}, \mathcal{I}}$. Of course, $\mathcal{T}^{\mathbf{L},q}_{\mathbf{V}, \mathcal{I}}=\mathcal{T}^{\mathbf{L},q}_{\mathbf{V}}$  if $\mathcal{I}$ is empty.

Note that Section \ref{expurgation} provides an introductory example of application of these notations, in the special case of $\mathbf{L}=(l,l)$ and $\hat{\mathbf{x}}=\mathbf{x}_1=\mathbf{x}$.

\section{Expurgation} \label{expurgation}
\begin{figure}[H]
	\begin{center}
    \includegraphics[width=70mm]{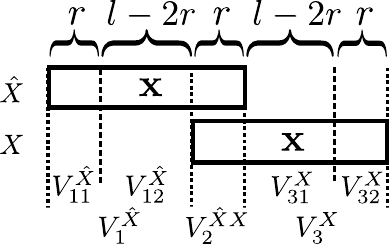}
    \caption{Further division of subblocks}
		\label{expurgationcikk2b}
		\end{center}
\end{figure}

\begin{Def}
A sequence $\mathbf{x} \in \mathcal{X}^{l}$ will be called $\gamma$-independent if its initial and final parts of length $r$ have empirical mutual information less than $\gamma$, for each $(\log l)^2 \le r \le \frac{l}{2}$. The subset of $\mathcal{T}^{l}_{P}$ consisting of $\gamma$-independent sequences is denoted by $\mathcal{T}^{l}_{P} (\gamma)$.
\end{Def}
\begin{Lem} \label{expurgationlemma}
For each $P \in \mathcal{P}^{l} (\mathcal{X})$ and $\gamma>0$
\begin{equation}
|\mathcal{T}_P^l \setminus \mathcal{T}_P^l(\gamma)| \le p(l) 2^{- (\log l)^2 \gamma} |\mathcal{T}_P^l|,
\end{equation}
where $p(l)$ denotes a polynomial factor not depending on $\gamma$.
\end{Lem}
\begin{figure}[H]
\begin{center}
\includegraphics[width=70mm]{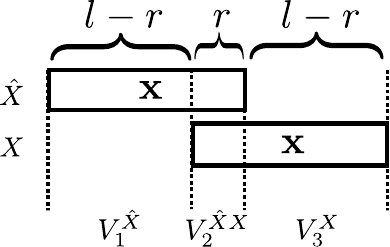}
\caption{$(\mathbf{L},q)$ array with $\mathbf{L}=(l,l)$, $q=l-r$ and $(\hat{\mathbf{x}},\mathbf{x}_1)=(\mathbf{x},\mathbf{x})$.}
\label{expurgationcikk2}
\end{center}
\end{figure}
\begin{IEEEproof}
We present a proof using the concept of $(\mathbf{L},q)$-array introduced in Section \ref{notation} as a simple example of the counting technique crucial for this paper. For a sequence $\mathbf{x}=x_1 \dots x_l$, the empirical mutual information of its initial and final parts of length $r$ is $\I_{V_2}(X \wedge \hat{X})$, where $(V_1,V_2,V_3)$ is the subtype sequence of the $(\mathbf{L},q)$-array with $\mathbf{L}=(l,l)$, $q=l-r$ and $\hat{\mathbf{x}}=\mathbf{x}_1=\mathbf{x}$, see Fig. \ref{expurgationcikk2}. For a fixed $(\log l)^2\le r \le \frac{l}{2}$ let $\mathcal{V}_{\gamma}^{l,r}$ be the set of subtype sequences $\mathbf{V}=(V_1,V_2,V_3)$ compatible with $(\mathbf{L},l-r)$ for which there exists $\mathbf{x} \in \mathcal{T}^{l}_{P}$ with $\mathds 1^{\mathbf{L},l-r}_{\mathbf{V}}(\mathbf{x};\mathbf{x})=1$ and $\I_{V_2}(X \wedge \hat{X})\ge \gamma$. Then
\begin{equation} \label{expurgeq1}
|\mathcal{T}_P^l \setminus \mathcal{T}_P^l(\gamma)| \le \sum_{r=(\log l)^2}^{\frac{l}{2}} \sum_{\mathbf{V} \in \mathcal{V}_{\gamma}^{l,r}}  |\mathcal{T}^{\mathbf{L},l-r}_{\mathbf{V}, \{\hat{\mathbf{x}}=\mathbf{x}_1\}}|,
\end{equation}
where $\mathcal{T}^{\mathbf{L},l-r}_{\mathbf{V}, \{\hat{\mathbf{x}}=\mathbf{x}_1\}}$ is defined as in the end of Section \ref{notation} with $\mathcal{I}$ consisting of the equality $\{\hat{\mathbf{x}}=\mathbf{x}_1\}$. We divide the first and last subblocks into two pieces according to Fig. \ref{expurgationcikk2b}. Formally for each $\mathbf{V}=(V_1,V_2,V_3)\in \mathcal{V}_{\gamma}^{l,r}$ we define
\begin{equation} \label{expurgset}
\mathcal{V}_{\gamma, \mathbf{V}}^{l,r} \triangleq \left\{\hspace{-3pt}
\begin{array}{l}
  (V_{11}^{\hat{X}},V_{12}^{\hat{X}},V_{31}^{X},V_{32}^{X}): \\
	V_{11}^{\hat{X}} \in \mathcal{P}^{r}(\mathcal{X}), V_{12}^{\hat{X}} \in \mathcal{P}^{l-2r}(\mathcal{X}), V_{11}^{\hat{X}}\oplus V_{12}^{\hat{X}}= V_{1}^{\hat{X}} \\
	V_{31}^{X} \in \mathcal{P}^{l-2r}(\mathcal{X}), V_{32}^{X} \in \mathcal{P}^{r}(\mathcal{X}), V_{31}^{X}\oplus V_{32}^{X}= V_{3}^{X} \\
	V_{11}^{\hat{X}}=V_2^{X}, V_{12}^{\hat{X}}=V_{31}^{X}, V_{32}^{X}= V_{2}^{\hat{X}}	
	\end{array}\hspace{-3pt}\right\}
\end{equation}
Then using (\ref{basicfact1}) and (\ref{basicfact2})
\begin{align}
&|\mathcal{T}^{\mathbf{L},l-r}_{\mathbf{V},  \{\hat{\mathbf{x}}=\mathbf{x}_1\}}|\le \sum_{(V_{11}^{\hat{X}},V_{12}^{\hat{X}},V_{31}^{X},V_{32}^{X}) \in \mathcal{V}_{\gamma, \mathbf{V}}^{l,r}} 2^{r \HH_{V_{2}} (\hat{X}X)} 2^{(l-2r)\HH_{V_{12}}(\hat{X})} \label{tovabboszt} \\
&=p'(l) 2^{r (\HH_{V_{2}} (\hat{X}X)-\HH_{V_2}(X)-\HH_{V_2}(\hat{X}))}
2^{(l-2r)\HH_{V_{12}}(\hat{X})+ r \HH_{V_{11}}(\hat{X})+r \HH_{V_{2}}(\hat{X})-l \HH_{P}(X)} 2^{l \HH_{P}(X)} \label{expuralgebra0} \\
&= p'(l)2^{-r\I_{V_2}(X \wedge \hat{X})} 2^{-l\J(V^{\hat{X}}_{11},V^{\hat{X}}_{12}, V^{\hat{X}}_{2})} 2^{l \HH_{P}(X)}, \label{expuralgebra2}
\end{align}
where $p'(l)$ denotes a polynomial factor not depending on $\gamma$. In (\ref{expuralgebra0}) we used that $V_{11}^{\hat{X}}=V_2^{X}$. Substituting (\ref{expuralgebra2}) into (\ref{expurgeq1}), the positivity of $J(V^{\hat{X}}_{11},V^{\hat{X}}_{12}, V^{\hat{X}}_{2})$, (\ref{basicfact1}) and the fact that $\I_{V_2}(X \wedge \hat{X})\ge \gamma$ prove the lemma.
\end{IEEEproof}

We will need the following consequence of Lemma \ref{expurgationlemma}: for any positive numbers  $\gamma_l$ with $\gamma_l \log l \rightarrow \infty$ as $l \rightarrow \infty$, for $l$ large enough
\begin{equation} \label{expurgsize}
\frac{1}{2} |\mathcal{T}^{l}_{P}| \le |\mathcal{T}^{l}_{P} (\gamma_l)| \le |\mathcal{T}^{l}_{P}|
\end{equation}

\section{The model}\label{Modell}

The transmitter has a codebook library with multiple constant composition codebooks. The codewords' length and type are fixed within codebooks, but can vary from codebook to codebook, subject to a bound on permissible codeword length ratios.

\begin{Def} \label{constantcomposition}
Let $D \in (0,1]$, positive integers $n$, $M$, $l^1,l^2,\dots,l^M$ with $D n \le l^i\le n$ for all $i \in [M]$, distributions $\{ P^{i} \in \iP^{l^i} (\iX), {i}\in [M] \}$ and rates $\{ R^{i} , i \in [M]\}$ be given parameters. A codebook library with the above parameters, denoted by $\mathcal{A}$, consists of constant composition codebooks $(A^{1},\dots,A^{M})$ such that $A^{i}=\{\vx^{i}_1,\vx^{i}_2, \dots \vx^{i}_{N^i}  \} $ with $\vx^{i}_{a} \in \mathcal{T}^{l^i}_{P^{i}}$, ${i} \in [M]$, $N^i = \left\lfloor 2^{l^{i} R^i} \right\rfloor$, $a \in [N^i]$. In the sequel, $n$ will be referred to as length-bound.
\end{Def}

The parameters in Definition \ref{constantcomposition} will depend on $n$, except for the constant $D$, but this dependence will be suppressed for brevity. Actually the proof of Theorem \ref{maintheorem} works if $D=D(n)$ goes to $0$ appropriately slowly. Note, however, that the appropriate speed of its convergence to $0$ would depend on the number of codebooks $M(n)$. For this reason and for the sake of simplicity we have chosen to fix $D$.

\begin{figure}[H]
\begin{center}
\includegraphics[width=70mm]{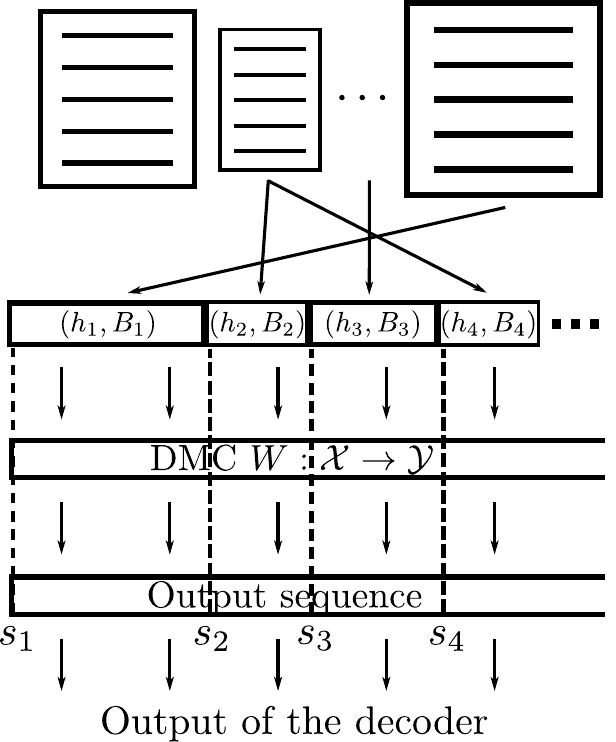}
\caption{Outline of the model}
\label{librarybolmodellcikk}
\end{center}
\end{figure}

The transmitter continuously sends messages to the receiver through a DMC $W: \mathcal{X} \rightarrow \mathcal{Y}$ that may be unknown to the sender and receiver. Before sending a message, the transmitter arbitrarily chooses one codebook of the library. This choice is not known to the receiver, who is cognisant only of the codebook library. His choices are described by an infinite codebook index sequence $\mathbf{h}=(h_1,h_2, h_3, \dots, h_j, \dots)$ where $h_j \in [M]$. In the sequel, $\mathbf{h}$ will be referred to as codebook schedule. To each fixed codebook schedule $\mathbf{h}$ there corresponds a sequence $B_1$, $B_2$, \dots of mutually independent random messages, where $B_j$ is uniformly distributed on $[N^{h_j}]$. To transmit $B_j=b$, the encoder assigns to it the $b$'th codeword of the codebook of index $h_j$. The transmission of this message starts at instance $s_j=\sum_{i=1}^{j-1} l^{h_i}+1$, depending on the codebook schedule but not on the actual messages.

\begin{Rem}
All formal statements in this paper refer to a fixed codebook schedule. Still, Theorem \ref{maintheorem} below covers also scenarios where the schedule $\mathbf{h}$ is random (random process of any kind), providing the messages $B_j$ are conditionally independent with uniform conditional distributions given $\mathbf{h}$, as the bounds (\ref{felsobecsles}) and (\ref{rcb}) do not involve $\mathbf{h}$.
\end{Rem}

A decoder is defined as a mapping of infinite channel output sequences  $\mathbf{y}=y_1,y_2,\dots$ into decoder output sequences $\mathbf{o}=o_1,o_2, \dots$, where each $o_t$ either equals a pair $(h,b)$ with $h \in [M]$, $b \in [N^{l^{\hat{h}}}]$, or the space symbol "-", or the string "erasure". By correct decoding of message $B_j=b$ we mean that $o_t$ is equal to $(B_j,b)$ at the starting instance $t=s_j$ and to space at the remaining $l^{h_j}-1$ instances of the transmission of this message. Accordingly, the average decoding error probability for the $j$'th transmission is
\begin{equation} \label{errordecoding}
Err_j^{\mathbf{h}} \triangleq Pr\left( O_{s_j} \ne (h_j,B_j) \text{ or } \exists \text{ } i \in \{ s_j+1,\dots,s_{j}+l^{h_j}-1 \}  \text{ with } O_{i} \ne "-"   \right).
\end{equation}
As correct decoding with small error probability is not possible for codebooks with type $P$ and rate $R\ge I(P,W)$, when such codebook is used a good decoder should declare erasure. We define average erasure detection failure probability for the $j$'th message as
\begin{equation} \label{errorth}
Edf_{j}^{\mathbf{h}} \triangleq Pr\left( \exists \text{ }  i \in \{ s_j,\dots,s_{j}+l^{h_j}-1 \} \text{ with } O_{i} \ne \text{ "erasure" }  \right).
\end{equation}
It is required to be small when $R^{h_j}\ge I(P^{h_j},W)$.

Note that in (\ref{errordecoding}) and (\ref{errorth}) the probabilities are calculated over the random choice of the messages, and the channel transitions. Capital letters are used to indicate randomness.

\begin{Rem}
The universal decoder in the proof of Theorem \ref{maintheorem} does not use the whole channel output sequence to determine $o_t$ but only $y_t$ and the preceding $2\cdot l^{max}-2$ and the subsequent $2\cdot l^{max}-2$ output symbols, where $l^{max}$ is the maximal codeword length in the codebook library, i.e., $l^{max}=\max_{i \in [M]} l^i \le n$. This can be seen to imply that the error events in (\ref{errordecoding}) (or in (\ref{errorth})) corresponding to message transmission indices $j_1$, $j_2$ are independent  if $|j_1 - j_2|$ exceeds a constant times $\frac{1}{D}$.
\end{Rem}
\begin{Rem}
The fact that there is only one sender raises the question whether it is possible to substitute average error terms (\ref{errordecoding}) and (\ref{errorth}) for maximal ones in Theorem \ref{maintheorem} below, as for example in \cite{Csiszar}. Unfortunately, as the error events depend simultaneously on several codebooks, this does not seem possible. The standard argument for upgrading average error results to maximal error gives only that the statement of Theorem \ref{maintheorem} also holds for the following error terms:
\begin{align} \label{maxerror}
&Errm_{j}^{\mathbf{h}} \triangleq \max_{b \in [N^{h_j}]} Pr\left( O_{s_j} \ne (h_j,B_j) \text{ or } \exists \text{ }  i \in \{ s_j+1,\dots,s_{j}+l^{h_j}-1 \}  \text{ with } O_{i} \ne "-"  | B_j=b \right),\\
&Edfm_{j}^{\mathbf{h}} \triangleq \max_{b \in [N^{h_j}]} Pr\left( \exists i \text{ }  \in \{ s_j,\dots,s_{j}+l^{h_j}-1 \} \text{ with } O_{i} \ne \text{ "erasure" } | B_j=b \right). \label{maxerrorc}
\end{align}
Here the maximum is taken relative to the $j$'th transmission, while still averaging relative to the other transmissions.
\end{Rem}

\section{Main theorem} \label{maintheoremsection}

\begin{Thm} \label{maintheorem}
For each $n$ let codebook library parameters as in Definition \ref{constantcomposition} be given with $D$ fixed and $\frac{1}{n}log M \rightarrow 0$ as $n \rightarrow \infty$. Then there exist a sequence $\nu_n(|\mathcal{X}|,|\mathcal{Y}|,\{M\}_{n=1}^{\infty},D)$ with $\frac{1}{n}\log \nu_n \rightarrow 0$ and for each $n$ a codebook library $\mathcal{A}$ with the given parameters, and decoder mappings such that for all codebook schedule $\mathbf{h}$ and index $j$
\begin{enumerate}[(i)]
\item \begin{equation} \label{felsobecsles}
Err_j^{\mathbf{h}} \le \nu_n \cdot 2^{- l^{h_j} \mathcal{E}_{r}(R^{h_j},P^{h_j},W)  },
\end{equation}
where
\begin{equation} \label{rcb}
\mathcal{E}_{r}(R,P,W) =\min_{\sumfrac{V\in \mathcal{P}(\mathcal{X} \times \mathcal{Y})}{V_{X}=P}} \D(V_{Y|X}||W|P)+|\I_V (X \wedge Y)-R|^{+}
\end{equation}
is the random coding error exponent function.
\item If $R^{h_j}\ge I[W,P^{h_j}]$ then
\begin{equation} \label{felsobecsles2}
Edf_{j}^{\mathbf{h}} \le \frac{1}{n}\log \nu_n.
\end{equation}
\end{enumerate}
\end{Thm}
\begin{Rem}
As the random coding exponent function $\mathcal{E}_{r}(R,P,W)$ is positive if and only if $R<I(P,W)$ , the bound (\ref{felsobecsles}) can be useful only if $R^{h_j}<I(P^{h_j},W)$. Recall that all parameters in Theorem \ref{maintheorem} depend on length-bound $n$. Even if $R^{h_j}<I(P^{h_j},W)$, the first factor in (\ref{felsobecsles}) may override the second one, but this does not happen for large $n$ if $I(P^{h_j},W)-R^{h_j}$ is bounded away from $0$. Then (\ref{felsobecsles}) guarantees exponentially small error probability, with exponent $\mathcal{E}_{r}(R^{h_j},P^{h_j},W)$ relative to codeword length $l^{h_j}$. This result is the best possible for codebooks whose rates $R^{h_j}$ are sufficiently close to $I(P^{h_j},W)$, since even for a single codebook with codeword type $P$ the random coding error exponent is tight for rates less than $I(P,W)$ but larger than a critical rate $\tilde{R}(P,W)$. Possible improvements for codebooks of small rates are beyond the scope of this paper. For the standard mentioned properties of the function $\mathcal{E}_{r}(R,P,W)$ see for example \cite{Csiszar2}.
\end{Rem}
\begin{Rem} \label{etan}
For erasure declaration failure probability, Theorem \ref{maintheorem} asserts only convergence to $0$, perhaps not exponentially fast. An argument similar to \cite{isit2013}, Appendix C suggests that this may not be a shortcoming due to loose calculation, upper bound (\ref{hibabindikator2th}) in the proof of Theorem \ref{maintheorem} is not exponentially small, under reasonable assumptions. An exponentially small erasure declaration failure probability could be achieved by modifying the decoder used in Theorem \ref{maintheorem}, replacing the threshold $\eta_n \rightarrow 0$ in (\ref{dekodolomukth}) by a positive contant, but at the expense of decreasing the decoding error probability exponents and perhaps declaring erasure also when decoding would be possible. As shown in Section \ref{improvements}, this problem, however, can be easily overcome if the sender knows the channel.
\end{Rem}

The next packing lemma provides the appropriate codebook library for Theorem \ref{maintheorem}. We emphasize that the constructed codebook library works simultaneously for all codebook schedules $\mathbf{h}$.

\begin{figure}[h]
\begin{center}
\includegraphics[scale=1.5]{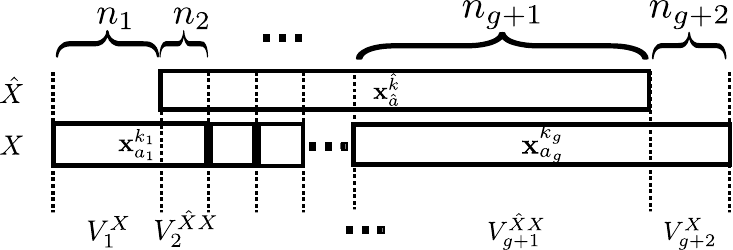}
\caption{Notations used in Lemma \ref{Rc-Packing-lemma}}
\label{packingtenycikk}
\end{center}
\end{figure}

Given codebook library parameters as in Def. \ref{constantcomposition}, for any sequence $\mathbf{k}=(\hat{k}, k_1,\dots, k_g)$ consisting of codebook indices (integers in $[M]$) let $\mathbf{L}(\mathbf{k})$ denote the sequence $(l^{\hat{k}},l^{k_1},\dots,l^{k_g})$. Further, given also a nonnegative integer $q$ satisfying (\ref{windowcondition1}) and (\ref{windowcondition2}) below, denote by $\mathcal{V}^{\mathbf{k},q,n}$ the family of those subtype sequences $\mathbf{V}=(V_1,... V_{g+1})$ compatible with $(\mathbf{L}(\mathbf{k}),q)$ for which the sequences $\hat{\mathbf{x}}$ and $\mathbf{x}_1,\dots,\mathbf{x}_g$ that form $(\mathbf{L}(\mathbf{k}),q)$-arrays with subtype sequence $\mathbf{V}$ have types $P^{\hat{k}}$ and $P^{k_i}$, ${i\in [g]}$ respectively.

\begin{Lem}\label{Rc-Packing-lemma}
Let a sequence of codebook library parameters be given as in Theorem \ref{maintheorem}. Then there exist a sequence $\nu^{'}_n (|\mathcal{X}|,\{M\}_{n=1}^{\infty},D)$ with $\frac{1}{n} \log \nu'_n \rightarrow 0$ and for each $n$ a codebook-library $\mathcal{A}$ with the given parameters such that each codeword is $\gamma_n$-independent, i.e., $\vx^{i}_{a} \in \mathcal{T}^{l^i}_{P^{i}}(\gamma_n)$, ${i} \in [M]$, $a \in [N^i]$ where $\gamma_n=(\log n)^{-\frac{1}{2}}$, and for each sequence $\mathbf{k}=(\hat{k}, k_1,\dots, k_g)$ consisting of codebook indices, non-negative integer $q$ with
\begin{align}
&q < l^{k_g}, \text{ } \sum_{i=2}^{g} l^{k_i}-q < l^{\hat{k}}, \label{windowcondition1}\\
&l^{\hat{k}}\le \sum_{i=1}^{g} l^{k_i}-q, \label{windowcondition2}
\end{align}
and subtype sequence $\mathbf{V}=(V_1,V_2,\dots, V_{g+2}) \in \mathcal{V}^{\mathbf{k},q,n}$
\begin{align}
&K^{\mathbf{k},q}[\mathbf{V}] \triangleq \sideset{}{'}\sum_{\sumfrac{a_i \in [N^{k_i}], i \in [g]}{\hat{a} \in [N^{\hat{k}}]}}\hspace{-11pt}\mathds 1^{\mathbf{L}(\mathbf{k}), q}_{\mathbf{V}}(\mathbf{x}_{\hat{a}}^{\hat{k}};\vx_{a_1}^{k_1}, \dots, \vx_{a_g}^{k_g}) \label{packingegyutt}\\
&\le \nu^{'}_n \cdot 2^{-\sum_{i=2}^{g+1} n_i \I_{V_i}(X \wedge \hat X) + \sum_{i \in [g]} l^{k_i}R^{k_i}+l^{\hat{k}}R^{\hat{k}}-l^{\hat{k}} \J(V_2^{\hat{X}}, \dots, V_{g+1}^{\hat{X}}) }, \notag
\end{align}
where in (\ref{packingegyutt}) the summation sign with the comma denotes standard summation except in the case of $g=1$,$\hat{k}=k_1$,$q=0$ when it is restricted for $\hat{a}$ in $[N^{\hat{k}}]=[N^{k_1}]$ to $\hat{a}\ne a_1$.
\end{Lem}
Note that (\ref{windowcondition1}) corresponds to (\ref{szimmetrikusfeltetel}) ensuring that in an $(\mathbf{L}(\mathbf{k}),q)$-array $\hat{\mathbf{x}}$ has nonempty overlap with both $\mathbf{x}_1$ and $\mathbf{x}_g$ while (\ref{windowcondition2}) ensures that, in addition, the first row is completely covered by the second row (see Fig. \ref{packingtenycikk}). Moreover, the condition (\ref{windowcondition1}) implies that $g \le \frac{2}{D}+1$.

\begin{IEEEproof}
Choose the codebook library $\mathcal{A}$ at random, i. e., for all $i \in [M]$ the codewords of $\mathcal{A}^i$ are chosen independently and uniformly from $\mathcal{T}^{l^i}_{P^i} (\gamma_n)$. Fix arbitrarily a sequence $\mathbf{k}=(\hat{k}, k_1,\dots, k_g)$ consisting of codebook indices, a non-negative integer $q$ fulfilling (\ref{windowcondition1}) and (\ref{windowcondition2}), a subtype sequence $\mathbf{V}=(V_1,V_2,\dots, V_{g+2}) \in \mathcal{V}^{\mathbf{k},q,n}$ and codeword indices $\hat{a}\in [N^{\hat{k}}]$, $a_1 \in [N^{k_1}], \dots, a_g \in [N^{k_g}]$ such that if if $g=1$, $\hat{k}=k_1$, $q=0$ then $\hat{a}\ne a_1$. Then, as shown in Appendix \ref{bizonyitasvarhatoertek}, the following inequality holds
\begin{equation} \label{egysegesbecsles}
\mathds{E}\left( 1^{\mathbf{L},q}_{\mathbf{V}}(\mathbf{X}_{\hat{a}}^{\hat{k}};\mathbf{X}_{a_1}^{k_1}, \dots, \mathbf{X}_{a_g}^{k_g})\right)\le \nu^{''}_n  2^{-\sum_{i=2}^{g+1} n_i \I_{V_i}(X \wedge \hat X) -l^{\hat{k}} \J(V_2^{\hat{X}}, \dots V_{g+1}^{\hat{X}})},
\end{equation}
where $\nu^{''}_n$ is a subexponential function of $n$ that depends only on $D$ and the alphabet size $|\mathcal{X}|$. Let $E^{\mathbf{k},r}[\mathbf{V}]$ be the exponent in upper-bound (\ref{packingegyutt}), i.e.,
\begin{equation}
E^{\mathbf{k},q}[\mathbf{V}]=-\sum_{i=2}^{g+1} n_i \I_{V_i}(X \wedge \hat X) -l^{\hat{k}} \J(V_2^{\hat{X}}, \dots, V_{g+1}^{\hat{X}})+ \sum_{i \in [g]} l^{k_i}R^{k_i}+l^{\hat{k}}R^{\hat{k}}.
\end{equation}
It follows from (\ref{egysegesbecsles}) that under this random selection the expected value of the expression
\begin{equation} \label{eztbecsultuk}
K^{\mathbf{k},q}[\mathbf{V}]2^{-E^{\mathbf{k},q}[\mathbf{V}]}
\end{equation}
is upper-bounded by $\nu^{''}_n$.

Denote by $S$ the sum of (\ref{eztbecsultuk}) for all possible $\mathbf{k}=(\hat{k}, k_1,\dots, k_g)$, $q$ and subtype sequences $\mathbf{V}$. As $M$ grows at most subexponentially and the number of types is polynomial, it follows that $\mathds{E}(S) \le \nu'_{n}$ for suitable $\nu^{'}_n (|\mathcal{X}|,\{M\}_{n=1}^{\infty},D)$ with $\frac{1}{n} \log \nu'_n \rightarrow 0$. Then there exists a realization of the codebook library with $S \le \nu'_{n}$. Hence, the lemma is proved if (\ref{egysegesbecsles}) is proved.  \end{IEEEproof}

\begin{Rem}
We would like to emphasize some interesting features of the rather technical proof of (\ref{egysegesbecsles}) in Appendix \ref{bizonyitasvarhatoertek}: in subcases 1b and 1d  it exploits the $\gamma_n$-independence property of the codewords and in subcases 2c and 2d second order types are employed.
\end{Rem}

\begin{IEEEproof}[Proof of Theorem \ref{maintheorem}]
Lemma \ref{Rc-Packing-lemma} provides the appropriate codebook library $\mathcal{A}$. We define the following sequential decoder. Assume that decoding related to symbols $y_1,\dots, y_{t-1}$ is already performed (i.e., $o_1,o_2, \dots,o_{t-1}$ are already defined) and now instance $t$ is analyzed. In the first stage of decoding the decoder tries to find indices $\tilde{h}, \tilde{b}$ which uniquely maximize
\begin{equation} \label{dekodolomuk}
l^h\left( \I (\mathbf{x}_{b}^{h} \wedge y_t y_{t+1} \dots y_{t+l^h -1})-R^{l^h} \right).
\end{equation}
If the decoder successfully finds a unique maximizer $\tilde{h}, \tilde{b}$, the second stage of decoding starts.

Let $\eta_n=\eta_n(\mathcal{X},\mathcal{Y},\{M\}_{n=1}^{\infty},D)$ be a sequence with $\eta_n \rightarrow 0$ as $n \rightarrow \infty$. In the second stage if
\begin{equation} \label{dekodolomukth}
\left( \I (\mathbf{x}_{\tilde{b}}^{\tilde{h}} \wedge y_t y_{t+1} \dots y_{t+l^{\tilde{h}} -1})-R^{l^{\tilde{h}}} \right) > \eta_n
\end{equation}
and for all $h, b$ and $d \in \{t-l^h+1, \dots, t-1 \} \cup \{t+1, \dots, t+l^{\tilde{h}}-1 \}$ the maximum of (\ref{dekodolomuk}) is strictly larger than
\begin{equation} \label{dekodolomuk2}
l^h\left( \I (\mathbf{x}_{b}^{h} \wedge y_{d} y_{d+1}\dots y_{d+l^h -1})-R^{l^h} \right),
\end{equation}
the decoder decodes $\mathbf{x}_{\tilde{b}}^{\tilde{h}}$ as the codeword sent in the window $[t,t+l^{\tilde{h}}-1]$, i.e., $o_{t}$ becomes equal to $(\tilde{h}, \tilde{b})$, and $o_{t+i}$ becomes equal to $\text{"-"}$, $i \in [l^{\tilde{h}}-1]$. Then the decoder jumps to the instance $t+l^{\tilde{h}}$, where the same but shifted procedure is performed. If in the first stage the maximum is not unique or in the second stage at least one of the required inequalities is not fulfilled, the decoder reports erasure in instance $t$, i.e., $o_t$ becomes equal to $\text{"erasure"}$, and the decoder goes to instance $t+1$. See Fig. \ref{stage2}.

\begin{figure}[H]
\begin{center}
\includegraphics[width=95mm]{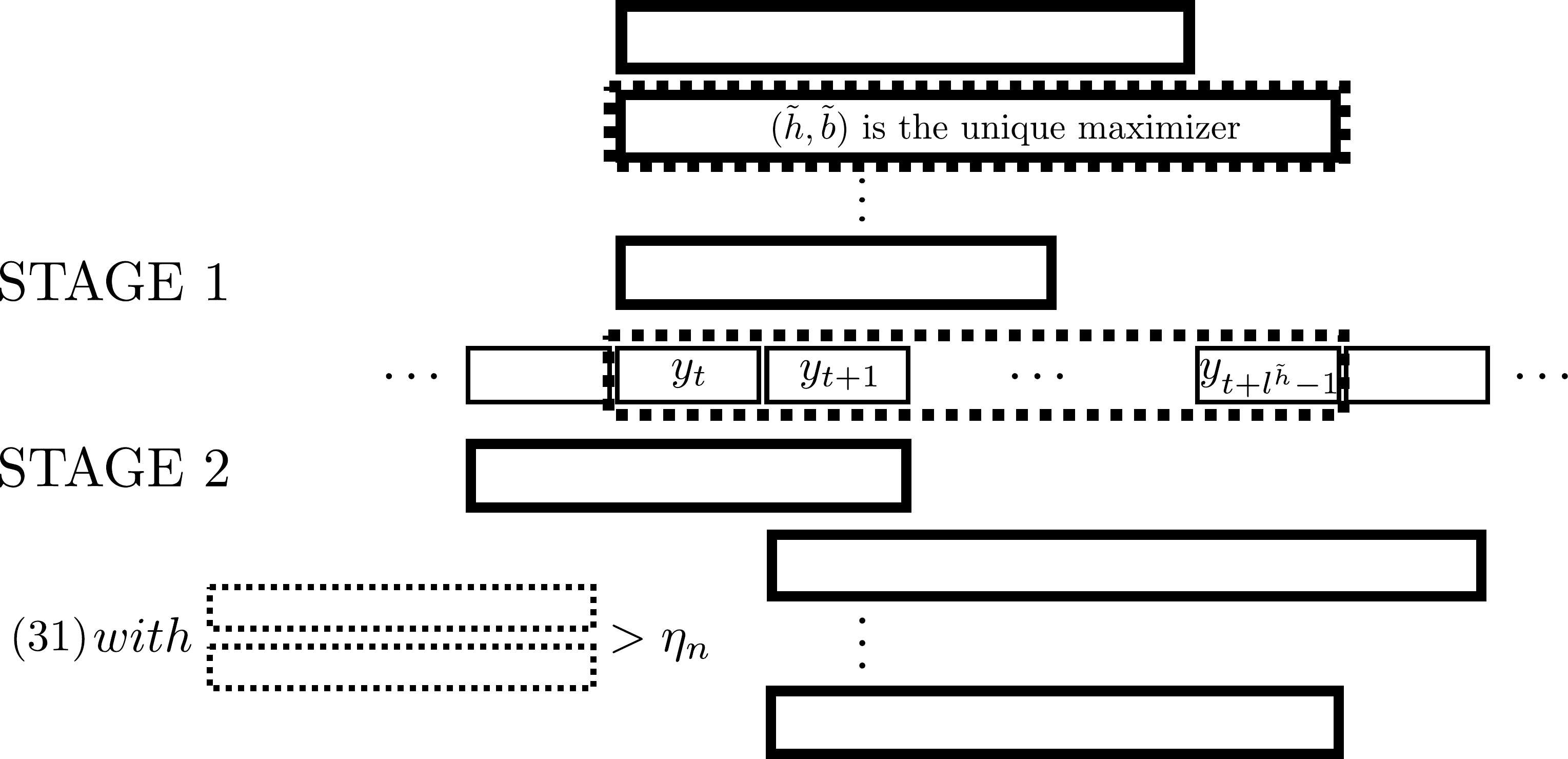}
\caption{Universal two-stage decoder}
\label{stage2}
\end{center}
\end{figure}

We prove that the codebook library $\mathcal{A}$ provided by Lemma \ref{Rc-Packing-lemma} with the decoder specified above with appropriately chosen $\eta_n(|\mathcal{X}|,|\mathcal{Y}|, \{M\}_{n=1}^{\infty},D)$ fulfills Theorem \ref{maintheorem}.

Let a codebook schedule $\mathbf{h}$ and an index $j$ be given. Let $Y_{s_j}, Y_{s_j+1},\dots, Y_{s_j+l^{h_j}-1}$ denote the random output symbols affected by the $j$-th message $B_j$.

\emph{Proof of part (i) of Theorem \ref{maintheorem}:}

Let $\mathcal{E}_j^{\mathbf{h}}(\text{TH})$ be the following event corresponding to threshold criterion (\ref{dekodolomukth})
\begin{equation}
\left\{ \left( \I (\mathbf{x}_{B_j}^{h_j} \wedge Y_{s_j}Y_{s_j+1} \dots Y_{s_j+l^{h_j}-1})-R^{l^{h_j}} \right)\le \eta_n \right\},
\label{errorpatternth}
\end{equation}
and for $\hat{k} \in [M]$ and $d \in \{s_j-l^{\hat{k}}+1,s_j-l^{\hat{k}}+2, \dots, s_j+l^{h_j}-1 \}$ let $\mathcal{E}_j^{\mathbf{h}}(\hat{k},d)$ be the event
\begin{equation}
\left\{\hspace{-3pt}\begin{array}{l}
l^{\hat{k}}\left( \I (\mathbf{x}_{\hat{a}}^{\hat{k}} \wedge Y_{d} Y_{d+1}\dots Y_{d+l^{\hat{k}} -1})-R^{l^{\hat{k}}} \right) \\
\ge l^{h_j}\left( \I (\mathbf{x}_{B_j}^{h_j} \wedge Y_{s_j} Y_{s_j+1}\dots Y_{s_j+l^{h_j}-1})-R^{l^{h_j}} \right), \\
\text{for some } \hat{a} \in [N^{\hat{k}}] \text{ }(\hat{a}\ne B_j \text{ if } \hat{k}=h_j \text{ and } d=s_j)
\end{array}\hspace{-3pt}\right\}. \label{errorpattern}
\end{equation}
The mutual informations in (\ref{errorpatternth}) and (\ref{errorpattern}) are empirical ones as in (\ref{dekodolomukth}), (\ref{dekodolomuk2}), though involving random sequences as the capital letters indicate. Denote by $Err_j^{\mathbf{h}}(\text{TH})$ and $Err_j^{\mathbf{h}}(\hat{k},d)$ the probabilities of these events, respectively. Then
\begin{equation} \label{uniokorlatahibara}
Err_j^{\mathbf{h}} \le Err_j^{\mathbf{h}}(\text{TH}) + \sum_{(\hat{k},d)} Err_j^{\mathbf{h}}(\hat{k},d).
\end{equation}

Note that the first term in (\ref{uniokorlatahibara}) is the probability that the threshold criterion is not fulfilled by the sent codeword while the sum of terms $Err_j^{\mathbf{h}}(\hat{k},d)$ provide upper-bound to the probability of the event that the sent codeword is outperformed in terms of (\ref{dekodolomuk}) either in Stage 1 or 2. The event that the decoder skips time index $s_j$ due to an erroneous previous decoding, i.e., $O_{s_j}=\text{"-"}$, is contained in the latter event.


By standard argument it follows from (\ref{basicfact1})-(\ref{basicfact4}) that
\begin{equation}
Err_j^{\mathbf{h}}(\text{TH}) \le subexp(n) \cdot 2^{- l^{h_j} \mathcal{E}^n_{TH}(R^{h_j},P^{h_j},W)} \label{bizTH},
\end{equation}
where
\begin{equation} \label{rcbth}
\mathcal{E}^n_{TH}(R^{h_j},P^{h_j},W) \triangleq \min_{\sumfrac{V\in \mathcal{P}(\mathcal{X} \times \mathcal{Y})}{V_{X}=P^{h_j}, \I_V(X \wedge Y)-R^{h_j}\le \eta_n }} \D(V_{Y|X}||W |P^{h_j}).
\end{equation}
It follows from (\ref{rcb}) that $\mathcal{E}_{r}(R^{h_j},P^{h_j},W) \le \mathcal{E}^n_{TH}(R^{h_j},P^{h_j},W)+\eta_n$. Hence, as $\eta_n \rightarrow 0$ we get
\begin{equation}
Err_j^{\mathbf{h}}(\text{TH}) \le subexp(n) \cdot 2^{- l^{h_j} \mathcal{E}_{r}(R^{h_j},P^{h_j},W)}. \label{biz1}
\end{equation}
For $\hat{k}=h_j$ and $d=s_j$ by standard argument
\begin{equation}
Err_j^{\mathbf{h}}(\hat{k},d) \le subexp(n) \cdot 2^{- l^{h_j} \mathcal{E}_{r}(R^{h_j},P^{h_j},W)} \label{biz1b}
\end{equation}

\begin{figure}[h]
\begin{center}
\includegraphics[scale=1.5]{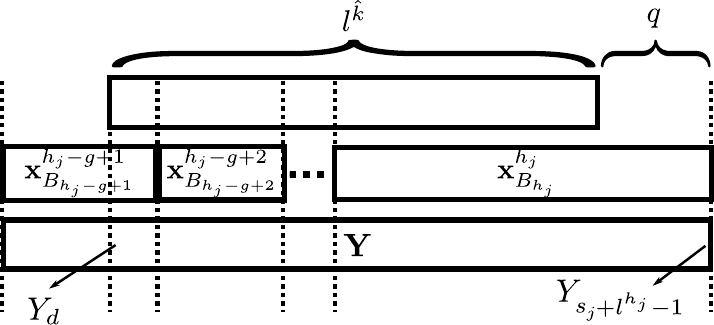}
\caption{Messages and assigned codewords affecting output sequence $\mathbf{Y}=(Y_{d} Y_{d+1}\dots Y_{s_j + l^{h_j}-1})$ along with explanation of notation $q$.}
\label{hibabecslesujcikkmod1}
\end{center}
\end{figure}

To prove part (i) of the theorem it is enough to show that this upper-bound also applies to $Err_j^{\mathbf{h}}(\hat{k},d)$ for all $\hat{k} \in [M]$ and $d \in \{s_j-l^{\hat{k}}+1,s_j-l^{\hat{k}}+2, \dots, s_j+l^{h_j}-1 \}$ (the number of pairs $(\hat{k},d)$ is subexponential). Fix a pair $(\hat{k},d) \ne (h_j,s_j)$. Assume that $d \le s_j$ (the analysis of the case $d>s_j$ is similar). The codebook schedule $\mathbf{h}$ determines the number $g$ of messages which affect outputs $Y_{d}, Y_{d+1},\dots, Y_{s_j + l^{h_j}-1}$ (see Fig. \ref{hibabecslesujcikkmod1}). Let
\begin{equation}
N \triangleq \prod_{i=1}^{g}  N^{h_{j-g+i}} = \prod_{i=1}^{g}2^{l^{h_{j-g+i}} R_{h_{j-g+i}}}.
\end{equation}
Then
\begin{equation}
Err_j^{\mathbf{h}}(\hat{k},d) = N^{-1} \sum_{a_i \in [N^{h_{j-g+i}}], i \in [g]}  \Prob \{ \mathcal{E}_j^{\mathbf{h}}(\hat{k},d) | B_{j-g+i}=a_i, i \in [g]\}. \label{teljesval}
\end{equation}

Let $k_i$ denote $h_{j-g+i}$, $i \in [g]$, let $\mathbf{k}=(\hat{k},k_1, \dots,k_g)$ and $\mathbf{L}(\mathbf{k})=(l^{\hat{k}}, l^{k_1}, \dots, k^{k_g})$. Furthermore, let $q= s_{j} + l^{h_j} - (d +  \hat{l})$ and $l=\sum_{i=1}^{g} l^{h_j - g +i}$ (see Fig. \ref{hibabecslesujcikkmod1}). This time we consider an array that includes the codewords $\vx_{\hat{a}}^{\hat{k}}, \vx_{a_1}^{k_1},\dots,\vx_{a_g}^{k_g}$ and also the output sequence $\vy \in \mathcal{Y}^l$, namely, we arrange $(\vx_{\hat{a}}^{\hat{k}}; \vx_{a_1}^{k_1},\dots,\vx_{a_g}^{k_g})$ into $(\mathbf{L}(\mathbf{k}),q)$-array as in Section \ref{notation} and we put the sequence $\mathbf{y}$ in a third row, the starting and ending position of $\mathbf{y}$ coincide with the starting position of $\vx_{a_1}^{k_1}$ and the ending position of $\vx_{a_g}^{k_g}$ respecticely. This 3-row array is also divided into subblocks according to the starting and ending positions of codewords $\vx_{\hat{a}}^{\hat{k}}, \vx_{a_1}^{k_1},\dots,\vx_{a_g}^{k_g}$. We apply the notations and conventions of Section \ref{notation}, for example, the lengths of the subblocks are denoted by $n_1,n_2,\dots,n_{g+2}$ as before. See Fig. \ref{hibabecslesujcikkmod2}.

Let $\mathcal{VM}^{\mathbf{k},q,n}$ be the following family of 3-row array subtype sequences
\begin{equation} \label{hibatipus}
\left\{\hspace{-3pt}
\begin{array}{l}
  \mathbf{V}=(V_1,V_2,\dots,V_{g+2}): V_1=V^{XY}_1 \in \mathcal{P}^{n_1}(\mathcal{X} \times \mathcal{Y}), V_{g+2}=V^{XY}_{g+2} \in \mathcal{P}^{n_{g+2}}(\mathcal{X} \times \mathcal{Y})\\
	V_i= V^{\hat{X}XY}_i \in \mathcal{P}^{n_i}(\mathcal{X} \times \mathcal{X} \times \mathcal{Y}), 2\le i \le g+1
	\\ (V^{X}_1,V^{\hat{X}X}_2,\dots, V^{\hat{X}X}_{g+1},V^{X}_{g+2}) \in \mathcal{V}^{\mathbf{k},q,n}\\
	l^{k_{g}} (\I_{V_{g+1} \oplus V_{g+2}}(X \wedge Y)-R^{k_{g}}) \le l^{\hat{k}} ( \I_{V_2 \oplus V_3 \oplus \dots \oplus V_{g+1}}(\hat{X} \wedge Y)-R^{\hat{k}})
	\end{array}\hspace{-3pt}\right\},
\end{equation}
where $\mathcal{V}^{\mathbf{k},q,n}$ is defined immediately prior to Lemma \ref{Rc-Packing-lemma}.

\begin{figure}[h]
\begin{center}
\includegraphics[scale=1.5]{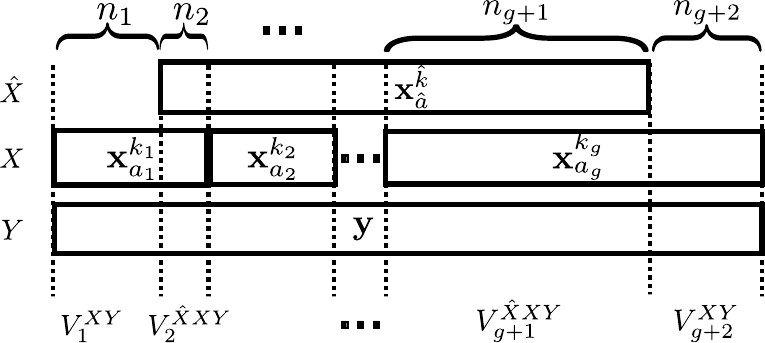}
\caption{The 3-row array of codewords $\hat{\mathbf{x}}, \mathbf{x}_1, \dots, \mathbf{x}_g$ and output $\mathbf{y}$.}
\label{hibabecslesujcikkmod2}
\end{center}
\end{figure}

Then, using (\ref{basicfact4}), the upper-bound (\ref{teljesval}) can be further upper-bounded by
\begin{align}
&\sum_{\mathbf{V} \in \mathcal{VM}^{\mathbf{k},q,n}} \hspace{-4pt} N^{-1} \hspace{-4pt} \sum_{a_i \in [N^{k_i}], i \in [g]} \hspace{-1pt} \prod_{i=1}^{g+2} \hspace{-1pt} \big( 2^{-n_i (\DD(V_i^{Y|X}||W|V_i^{X})+\HH_{V_i}(Y|X))}  \notag\\
&\cdot  \big| \{\mathbf{y} \in \mathcal{Y}^{l}:\mathds{1}^{\mathbf{L}(\mathbf{k}),q}_{\mathbf{V}} ( \vx_{\hat{a}}^{\hat{k}}; \vx_{a_1}^{k_1},\dots,\vx_{a_g}^{k_g};\vy) =1 \text{ for some } \hat{a} \in [N^{\hat{k}}] \big|, \label{hibabindikator}
\end{align}
where the indicator function $\mathds{1}^{\mathbf{L}(\mathbf{k}),q}_{\mathbf{V}} ( \vx_{\hat{a}}^{\hat{k}}; \vx_{a_1}^{k_1},\dots,\vx_{a_g}^{k_g};\vy)$ equals $1$ (otherwise $0$) if placing $\vx_{\hat{a}}^{\hat{k}}, \vx_{a_1}^{k_1},\dots,\vx_{a_g}^{k_g}$ and $\vy$ into a 3-row array as above, the type of the $i$'th subblock is $V_i$ for each $i \in [g+2]$ with $n_i>0$.

We can upper-bound the set size in (\ref{hibabindikator}) two different ways. The first bound is $2^{\sum_{i=1}^{g+2} n_i \HH_{V_i} (Y|X)}$. The second bound is:
\begin{equation}
\sum_{\hat{a} \in [N^{\hat{k}}]} 2^{n_1 \HH_{V_1} (Y|X) + \sum_{i=2}^{g+1} n_i \HH_{V_i} (Y|\hat{X}X)+n_{g+2} \HH_{V_{g+2}} (Y|X)} \mathds{1}^{\mathbf{L}(\mathbf{k}),q}_{\mathbf{V}'} ( \vx_{\hat{a}}^{\hat{k}}; \vx_{a_1}^{k_1},\dots,\vx_{a_g}^{k_g} ),
\end{equation}
where $\mathbf{V}' =(V_1^{X}, V_2^{\hat{X}X},\dots, V_{g+1}^{\hat{X}X},V_{g+2}^{X})$. Substituting these bounds into (\ref{hibabindikator}) and using (\ref{packingegyutt}) we get that:
\begin{align}
&Err_j^{\mathbf{h}}(\hat{k},d) \le \sup_{\mathbf{V} \in \mathcal{VM}^{\mathbf{k},q,n}} subexp(n) \hspace{-1pt} 2^{- \sum_{i=1}^{g+2} n_i \DD(V_i^{Y|X}||W|V_i^{X})} 2^{-|\sum_{i=2}^{g+1} n_i \I_{V_i}(\hat{X}\wedge X Y)+l^{\hat{k}}J(V_2^{\hat{X}}, \dots, V_{f+1}^{\hat{X}})-l^{\hat{k}}R^{\hat{k}}|^{+}} \label{hibabindikatorutolag} \\
&\le \sup_{\mathbf{V} \in \mathcal{VM}^{\mathbf{k},q,n}} subexp(n) \hspace{-1pt} 2^{- \sum_{i=1}^{g+2} n_i \DD(V_i^{Y|X}||W|V_i^{X})} 2^{-|\sum_{i=2}^{g+1} n_i \I_{V_i}(\hat{X}\wedge Y)+l^{\hat{k}}J(V_2^{\hat{X}}, \dots, V_{f+1}^{\hat{X}})-l^{\hat{k}}R^{\hat{k}}|^{+}}. \label{hibabindikator2}
\end{align}
The term inside $||^{+}$ is equal to
\begin{align}
-\sum_{i=2}^{g+1} n_i \HH_{V_i}(\hat{X} | Y) + l^{\hat{k}} \HH(P^{\hat{k}}) -l^{\hat{k}}R^{\hat{k}} \label{lassanuccso}
\end{align}
By convexity, (\ref{lassanuccso}) can be lower-bounded by
\begin{align}
&-l^{\hat{k}} \HH_{V_2 \oplus \dots \oplus V_{g+1}}(\hat{X} | Y) + l^{\hat{k}}\HH(P^{\hat{k}}) -l^{\hat{k}}R^{\hat{k}}=l^{\hat{k}} \I_{V_2 \oplus \dots \oplus V_{g+1}}(\hat{X}\wedge Y) -l^{\hat{k}}R^{\hat{k}}. \label{lassanuccso2}
\end{align}
Substituting (\ref{lassanuccso2}) into (\ref{hibabindikator2}) and using (\ref{hibatipus}) we get
\begin{align}
&Err_j^{\mathbf{h}}(\hat{k},d) \le  \sup_{\mathbf{V} \in \mathcal{VM}^{\mathbf{k},q,n}} subexp(n) \cdot 2^{-\sum\limits_{i=1}^{g+2} n_i \DD(V_i^{Y|X}||W|V_i^{X})-|l^{k_g} \I_{V_{g+1} \oplus V_{g+2}}(X \wedge Y) -l^{k_{g}} R^{k_g}|^{+}} \label{lassanuccso3} \\
&\le \sup_{\mathbf{V} \in \mathcal{VM}^{\mathbf{k},q,n}} subexp(n) \cdot 2^{-\sum\limits_{i=g+1}^{g+2} n_i \DD(V_i^{Y|X}||W|V_i^{X})-|l^{k_g} \I_{V_{g+1} \oplus V_{g+2}}(X \wedge Y) -l^{k_{g}} R^{k_g}|^{+}} \label{lassanuccso4}
\end{align}
Hence, the convexity of the divergence proves
\begin{equation}
Err_j^{\mathbf{h}}(\hat{k},d) \le subexp(n) \cdot 2^{- l^{h_j} \mathcal{E}_{r}(R^{h_j},P^{h_j},W)}.
\end{equation}

Note that in this part $\eta_n$ can be arbitrary positive sequence which goes to $0$ as $n\rightarrow \infty$. However, it will turn out from the proof of part (ii) that the sequence $\eta_n$ has to converge to $0$ sufficiently slowly.

\emph{Proof of part (ii) of Theorem \ref{maintheorem}:}

For $\hat{k} \in [M]$ and $d \in \{s_j-l^{\hat{k}}+1,s_j-l^{\hat{k}}+2, \dots, s_j+l^{h_j}-1 \}$ let $\mathcal{E}df_j^{\mathbf{h}}(\hat{k},d)$ be the event
\begin{equation}
\left\{\hspace{-3pt}\begin{array}{l}
\left( \I (\mathbf{x}_{\hat{a}}^{\hat{k}} \wedge Y_{d} Y_{d+1}\dots Y_{d+l^{\hat{k}} -1})-R^{l^{\hat{k}}} \right) > \eta_n \\
\text{for some } \hat{a} \in [N^{\hat{k}}] \text{ }(\hat{a}\ne B_j \text{ if } \hat{k}=h_j \text{ and } d=s_j)
\end{array}\hspace{-3pt}\right\}, \label{errorpatternii}
\end{equation}
and let $Edf_{j}^{\mathbf{h}}(\hat{k},d)$ denote its probability. Introduce also the notation $Edf_{j}^{\mathbf{h}}(\text{TH})=1-Err_j^{\mathbf{h}}(\text{TH})$. Then
\begin{equation} \label{uniokorlatahibaraii}
Edf_{j}^{\mathbf{h}} \le Edf_{j}^{\mathbf{h}}(\text{TH}) + \sum_{(\hat{k},d)} Edf_{j}^{\mathbf{h}}(\hat{k},d).
\end{equation}
By standard argument it follows from (\ref{basicfact2}) and (\ref{basicfact4}) that
\begin{equation}
Edf_{j}^{\mathbf{h}}(\text{TH}) \le subexp(n) \cdot 2^{- l^{h_j} \mathcal{E}df_{TH}^n (R^{h_j},P^{h_j},W)} \label{bizTHii},
\end{equation}
where
\begin{equation} \label{rcbthii}
\mathcal{E}df_{TH}^n (R^{h_j},P^{h_j},W) \triangleq \min_{\sumfrac{V\in \mathcal{P}(\mathcal{X} \times \mathcal{Y})}{V_{X}=P^{h_j}, \I_V(X \wedge Y)-R^{h_j} > \eta_n }} \D(V_{Y|X}||W|P^{h_j})=\min_{\sumfrac{V\in \mathcal{P}(\mathcal{X} \times \mathcal{Y})}{V_{X}=P^{h_j}, \I_V(X \wedge Y)-R^{h_j} > \eta_n }} \D(V||P^{h_j}W).
\end{equation}
In part (ii) it is assumed that $R^{h_j}\ge I[W,P^{h_j}]$. It follows that for each possible $V$ in (\ref{rcbthii})
\begin{equation} \label{rcbthii2}
\I_{V}(X \wedge Y)-I[W,P^{h_j}] > \eta_n.
\end{equation}
Hence using Lemma 2.7 of \cite{Csiszar2} it follows that there exists $\zeta_n (\eta_n, |\mathcal{X}|, |\mathcal{Y}|)$>0 such that for each possible $V$ in (\ref{rcbthii}) its variational distance from $P^{h_j}W$ is at least $\zeta_n$. Then (\ref{bizTHii}),(\ref{rcbthii}) and Pinsker inequality show that $Edf_{j}^{\mathbf{h}}(\text{TH})$ goes to $0$ if $\eta_n(|\mathcal{X}|,|\mathcal{Y}|, \{M\}_{n=1}^{\infty},D)$ converges to $0$ sufficiently slowly.

To prove part (ii) of the theorem it remains to show that also $ \sum_{(\hat{k},d)} Edf_{j}^{\mathbf{h}}(\hat{k},d)$ goes to $0$ if $\eta_n$ converges to $0$ sufficiently slowly. Fix a pair $(\hat{k},d)$. Assume that $d \le s_j$ (the analysis of the case $d>s_j$ is similar). Replicating the notations and arguments of the proof of part (i) of the theorem the analogue of (\ref{hibabindikator2}) with (\ref{lassanuccso}) and (\ref{lassanuccso2}) gives that
\begin{align}
&Edf_{j}^{\mathbf{h}}(\hat{k},d) \le \sup_{\mathbf{V} \in \mathcal{VM}_{edf}^{\mathbf{k},q,n}} subexp(n) \prod_{i=1}^{g+2} \hspace{-1pt} 2^{-n_i \DD(V_i^{Y|X}||W|V_i^{X})} 2^{-|l^{\hat{k}} \I_{V_2 \oplus \dots \oplus V_{g+1}}(\hat{X}\wedge Y) -l^{\hat{k}}R^{\hat{k}}|^{+}}, \label{hibabindikator2th}
\end{align}
where the family of subtype sequences $\mathcal{VM}_{edf}^{\mathbf{k},r,n}$ is equal to
\begin{equation} \label{hibatipusth}
\left\{\hspace{-3pt}
\begin{array}{l}
  \mathbf{V}=(V_1,V_2,\dots,V_{g+2}): V_1=V^{XY}_1 \in \mathcal{P}^{n_1}(\mathcal{X} \times \mathcal{Y}), V_{g+2}=V^{XY}_{g+2} \in \mathcal{P}^{n_{g+2}}(\mathcal{X} \times \mathcal{Y})\\
	V_i= V^{\hat{X}XY}_i \in \mathcal{P}^{n_i}(\mathcal{X} \times \mathcal{X} \times \mathcal{Y}), 2\le i \le g+1
	\\ (V^{X}_1,V^{\hat{X}X}_2,\dots, V^{\hat{X}X}_{g+1},V^{X}_{g+2}) \in \mathcal{V}^{\mathbf{k},q,n}, \I_{V_2 \oplus V_3 \oplus \dots \oplus V_{g+1}}(\hat{X} \wedge Y)-R^{\hat{k}} > \eta_n.
	\end{array}\hspace{-3pt}\right\}.
\end{equation}
From (\ref{hibabindikator2th}) and (\ref{hibatipusth}) it follows that $\sum_{(\hat{k},d)} Edf_{j}^{\mathbf{h}}(\hat{k},d)$ can be upper-bounded by $subexp(n) 2^{-l^{\hat{k}} \eta_n}$ which goes to $0$ if $\eta_n(|\mathcal{X}|,|\mathcal{Y}|, \{M\}_{n=1}^{\infty},D)$ converges to $0$ sufficiently slowly.
\end{IEEEproof}

\section{Improvement when the channel is known by the sender} \label{improvements}

Theorem \ref{maintheorem} provides a universal result: Neither the design of the codebook library nor the decoder depends on the channel matrix $W$. In this section we outline a substantial improvement when the channel is known to the sender (it remains unknown to the receiver).

Intuitively, supposing there are $M$ kinds of messages, with given rate and codeword length for each of them, the improvement will be based on an extended codebook library that contains several codebooks with different types for each message kind. From the codebooks available for a given message kind, the sender will chose the one whose type maximizes the random coding exponent for the actual channel, or if that maximum is $0$, the sender will use a one-codeword codebook just indicating the message kind. 

Formally, let a sequence of codebook library parameters be given as in Theorem \ref{maintheorem}, except for the prescribed types, i.e., let $D \in (0,1]$ be fixed, and for each $n$ let $M$ with $\frac{1}{n} \log M \rightarrow 0$, $l^1,l^2,\dots,l^M$ with $D n \le l^i\le n$ for all $i \in [M]$ and rates $\{ R^{i} , i \in [M]\}$ be given. 

We construct a sequence of codebook library parameters as follows. $D$ remains unchanged. Instead of $i \in [M]$, the codebook indices will be triplets $(i,P,s)$, where $i \in [M]$,  $P \in P^{l^i}(\mathcal{X})$ and $s \in \{0,1\}$. Let the length, type and rate of the codebook indexed by triplet $(i,P,s)$ be equal to $l^{(i,P,s)}=l^i$, $P^{(i,P,s)}=P$ and $R^{(i,P,s)}=R^i \cdot s$, respectively. As the number of triplets $(i,P,s)$ remains subexponential in $n$ we can apply Theorem \ref{maintheorem} with this modified sequence of codebook library parameters. The codebook library provided by Theorem \ref{maintheorem} contains two codebooks corresponding to each pair $(l^i,R^i)$ and type $P$, one with the rate $R^i$ and one with rate $0$, i.e., consisting of only one codeword. This codebook library will be referred to as extended codebook library.

An infinite message kind schedule $\mathbf{k}=(k_1,\dots,k_j \dots)$, $k_j \in [M]$, specifies for each $j$ the kind of message $k_j$ to be transmitted at time $j$. Relying on the channel knowledge, the sender constructs a codebook schedule $\mathbf{h}(\mathbf{k},W)=(h_1(k_1,W),\dots,h_j(k_j,W),\dots)$ as follows.

For each $i \in [M]$ define
\begin{equation}
\mathcal{E}_r^i (R,W) \triangleq \max_{P \in P^{l^{i}}(\mathcal{X})} \mathcal{E}_r (P,R,W),
\end{equation}
where $\mathcal{E}_r (P,R,W)$ is defined in (\ref{rcb}). Note that maximization for all $P \in \mathcal{P} (\mathcal {X})$, rather than only for $l_i$-types, gives the standard random coding exponent $E_r(R,W)$ of the channel. Thus, when $n$ and hence $l_i > Dn$ is large, $E_r^i (R,W)$ differs only negligibly from $E_r( R,W)$. 

Let $P_1$ and $P_0$ be types for which $\mathcal{E}_r^{k_j} (R,W)=\mathcal{E}_r (P_1,R^{k_j},W)$ and $\mathcal{E}_r^{k_j} (0,W)=\mathcal{E}_r (P_0,0,W)$ respectively. Let $h_j(k_j,W)$ be equal to $(k_j,P_1,1)$ if $E_r^{k_j} (R^{k_j},W) > 0$ and $(k_j,P_0,0)$ otherwise. This means that in each transmission $j$, the sender uses the optimal input distribution with the given rate. Moreover, when reliable message detection is not possible with the given rate, a $0$-rate (and hence reliable) codebook is used. In accordance with this codebook schedule construction, we modify the decoder used in the proof of Theorem \ref{maintheorem}. If the output of the decoder is the only codeword in the codebook indexed by $(i,P,0)$ for some $i \in [M]$ and $P \in P^{l^i}(\mathcal{X})$ then the decoder reports "erasure" and supplements it with declaring that the receiver wanted to send $i$'th message kind but the channel is not supported it. Altogether, the next corollary follows from Theorem \ref{maintheorem} (i) (part (ii) of Theorem \ref{maintheorem} is not used).  
\begin{Cor} \label{cor1}
Let $\nu_n$ be the sequence specified by Theorem \ref{maintheorem}. For each infinite message kind schedule $\mathbf{k}=(k_1,\dots,k_j \dots)$, $k_j \in [M]$, and index $j$, using the extended codebook library with codebook schedule $\mathbf{h}(\mathbf{k},W)=(h_1(k_1,W),\dots,h_j(k_j,W),\dots)$ and the decoder specified above the followings hold.
\begin{enumerate}[(i)]
\item If $\mathcal{E}_r^{k^j} (R^{k^j},W)>0$, the probability of incorrectly decoding the $j$'th message is less than $\nu_n \cdot 2^{- l^{k_j} \mathcal{E}_{r}^{k_j}(R^{k_j},W) }$.
\item If $\mathcal{E}_r^{k^j} (R^{k^j},W)=0$, the decoder reports "erasure" and declares that the kind of the erased message is $k_j$, with probability at least $1-\nu_n \cdot 2^{- l^{k_j} \mathcal{E}_{r}^{k_j}(0,W) }$.
\end{enumerate}
\end{Cor}
\begin{Rem}
The construction of $\mathbf{h}(\mathbf{k},W)=(h_1(k_1,W),\dots,h_j(k_j,W),\dots)$ above is very specific. Actually, for each transmission $j$ the sender can decide whether the exponent $\mathcal{E}_r^{k_j} (R^{k_j},W)$ is sufficient or not. If it is not sufficient for his purposes, he can choose to use the corresponding $0$-rate codebook instead of actual message transmission.
\end{Rem}
\begin{Rem} \label{singlecodebook}
Even the special case of Corollary \ref{cor1} for the classical situation of transmitting messages of a single kind is of interest: If the sender but not the receiver knows the channel the random coding exponent of the actual channel (i.e., the random coding exponent maximized over input distribution) is achievable. In the literature, this fact is usually stated only when both sender and receiver know the channel. Note that this special case of the corollary also follows from \cite{Csiszar}, though not explicitly stated there. 
\end{Rem}

\section{Discussion}

A generalization of the DMC coding theorem of \cite{Csiszar} has been studied allowing not just the rate and the type but also the length of the codewords to vary across codebooks. This generalization could provide a theoretical background for practical scenarios when different coding strategies are used for sending different kind of messages (e.g. audio, data, video). It has been shown that in this scenario simultaneously for each codebook choice of each transmission, the same error exponent can be achieved as the random coding error exponent for the chosen codebook alone, supplemented with non-exponential erasure detection. This has been achieved with a completely universal construction: Neither the design of the codebook library nor the decoder depends on the channel.

When the channel is known to the sender, an improvement is given while maintaining the universality of the decoder. The improvement leads to exponent also for erasure declaration failure probability and shows that the maximum of the random coding error exponent over the possible input distributions is achievable for each message kind. The possible improvement via relaxing the universality of the decoder is not addressed in this paper. However, we note that in the model with equal codeword lengths, \cite{Csiszar4} shows that maximum likelihood decoding admits to achieve, individually for each codebook, also the expurgated error exponents that for small rates exceeds the random coding exponent. A similar result likely holds also for the model in this paper.

The difficulty in the analysis of the model in this paper comes from the fact that the different codeword lengths cause a certain asynchronism at the receiver, who should also estimate the boundaries of the codewords and avoid error propagation. The asynchronous nature of this model gives a natural connection to "strong asynchronism" in \cite{Tchamkarten}. In that model the sender has only one codebook, sends a message only once in an exponentially large (in the codeword length) time window, when the sender is idle a special dummy symbol denoted by $*$ is transmitted. The time of the message transmission is not known to the receiver. Tradeoff between the rate of the codebook and the exponent of the time window is investigated. The detailed investigation of the relation of this model to ours is beyond the scope of this paper. However, in order to arouse the reader's attention we note the following.
\begin{enumerate}

\item It is a natural idea to try to employ Theorem \ref{maintheorem} in its current form to the model of strong asynchronism via artificially introducing one-codeword codebooks consisting of dummy symbols $*$. One problem with this idea is that the random coding error exponent of these artificial codebooks is $0$. A better option is to employ artificial one-codeword codebooks (similarly as in Section \ref{improvements}) with positive random coding error exponent to model the idle periods of the sender. Then the exponentially small error probability in each transmission ensures that the decoder fails only with small probability even in an exponential large time window. Hence, this application of Theorem \ref{maintheorem} would lead to a meaningful model, nevertheless, it would differ from the original model of strong asynchronism.

\item Theorem 3 of \cite{Polyanskiy} shows that the achievable pairs (rate, time window exponent) can be also achieved with a universal decoder. This is, however, shown only under an error criterion which does not require exact synchronization. Hence, the event of detecting the right codeword in a wrong position, partially overlapping with the correct one is omitted in the error analysis. \cite{HongkongML} provides error exponent in the model of strong asynchronism using maximum likelihood decoder. Here, exact synchronization is required but no simple single-letter expression is obtained for the exponent. As our paper does handle the event of partial overlap, we think that the tools used here can be used to strengthen Theorem 3 of \cite{Polyanskiy} and the error exponent analysis in the model of strong asynchronism.

\end{enumerate}

\appendices

\section{Proof of inequality (\ref{egysegesbecsles})} \label{bizonyitasvarhatoertek}
In this section we suppose that a sequence of codebook library parameters as in Lemma \ref{Rc-Packing-lemma} is given. Let $\gamma_n=(\log n)^{-\frac{1}{2}}$. Choose the codebook library $\mathcal{A}$ at random, i. e., for all $i \in [M]$ the codewords of $\mathcal{A}^i$ are chosen independently and uniformly from $\mathcal{T}^{l^i}_{P^i} (\gamma_n)$. We prove rigorously that for arbitrarily sequence $\mathbf{k}=(\hat{k}, k_1,\dots, k_g)$ consisting of codebook indices, non-negative integer $q$ fulfilling (\ref{windowcondition1}) and (\ref{windowcondition2}), subtype sequence $\mathbf{V}=(V_1,V_2,\dots, V_{g+2}) \in \mathcal{V}^{\mathbf{k},q,n}$, and indices $\hat{a}\in [N^{\hat{k}}]$, $a_1 \in [N^{k_1}], \dots, a_g \in [N^{k_g}]$ supposing $\hat{a}\ne a_1 \text{ if $g=1$, $\hat{k}=k_1$ and $q=0$}$, the following inequality holds
\begin{equation} \label{egysegesbecsles2}
\mathds{E}\left( 1^{\mathbf{L},q}_{\mathbf{V}}(\mathbf{X}_{\hat{a}}^{\hat{k}};\mathbf{X}_{a_1}^{k_1}, \dots, \mathbf{X}_{a_g}^{k_g})\right)\le \nu^{''}_n  2^{-\sum_{i=2}^{g+1} n_i \I_{V_i}(X \wedge \hat X) -l^{\hat{k}} \J(V_2^{\hat{X}}, \dots V_{g+1}^{\hat{X}})},
\end{equation}
where $\nu^{''}_n$ is a subexponential function of $n$ that depends only on $D$ and the alphabet size $\mathcal{X}$.

The proof relies strongly on the notations introduced in Section \ref{notation}. We define a set of equalities $\mathcal{I}$ as follows: for all $i \in [g]$, equality $\hat{\mathbf{x}}=\mathbf{x_i}$ is in $\mathcal{I}$ iff $(\hat{k},\hat{a})=(k_i,a_i)$ and for all $i,j \in [g]$ equality $\mathbf{x_i}=\mathbf{x}_j$ is in $\mathcal{I}$ iff $(k_i,a_i)=(k_j,a_j)$. For notational convenience we also define a set $\mathcal{I}^{*}$ consisting of the positive integers $j \in [g]$ such that $(k_i,a_i) \ne (k_j,a_j)$ for all $i<j$. Note that $\mathcal{I}$ determines $\mathcal{I}^{*}$ but the reverse is not true.

To prove (\ref{egysegesbecsles2}) we separately investigate two cases: $n_2$ and $n_{g+1}$ are both less than or equal to $n-(\log n)^2$ (case 1) or at least one of them is larger than $n-(\log n)^2$ (case 2). Both can be larger then $n-(\log n)^2$ only in the case $g=1$ when $n_2=n_{g+1}$; in this case always $\mathcal{I}=\emptyset$, and the proof of subcase 1a below works. From now on, we assume that $g \ge 2$.

\emph{CASE 1}: By symmetry it can be assumed that $n_2 \le n_{g+1}$ (this assumption ensures that the counting from left to right works in all subcases). We have to separately investigate four subcases: $\{ \hat{\mathbf{x}}=\mathbf{x}_1 \} \notin \mathcal{I}$ , $\{ \hat{\mathbf{x}} =\mathbf{x}_g \} \notin \mathcal{I}$ (subcase 1a), $\{ \hat{\mathbf{x}}=\mathbf{x}_1 \} \in \mathcal{I}$ , $\{ \hat{\mathbf{x}} =\mathbf{x}_g \} \notin \mathcal{I}$ (subcase 1b) $\{ \hat{\mathbf{x}}=\mathbf{x}_1 \} \notin \mathcal{I}$ , $\{ \hat{\mathbf{x}} =\mathbf{x}_g \} \in \mathcal{I}$ (subcase 1c) and $\{ \hat{\mathbf{x}}=\mathbf{x}_1 \} \in \mathcal{I}$ , $\{ \hat{\mathbf{x}} =\mathbf{x}_g \} \in \mathcal{I}$ (subcase 1d).

\emph{SUBCASE 1a}: For the sake of clarity first we assume that  not only equalities $\hat{\mathbf{x}}=\mathbf{x}_1 , \hat{\mathbf{x}}=\mathbf{x}_g$ are not in $\mathcal{I}$ but $\mathcal{I}$ is empty. This assumption is relaxed in the second part of the discussion of this subcase. Then using (\ref{basicfact2}) and (\ref{expurgsize}) it follows that
\begin{equation} \label{becslesalap}
\mathds{E}\left( 1^{\mathbf{L},q}_{\mathbf{V}}(\mathbf{X}_{\hat{a}}^{\hat{k}};\mathbf{X}_{a_1}^{k_1}, \dots, \mathbf{X}_{a_g}^{k_g})\right) \le subexp(n)\cdot 2^{-l^{\hat{k}}\HH(P^{\hat{k}})-\sum_{i=1}^{g} l^{k_i} \HH(P^{k_i})}\cdot |\mathcal{T}^{\mathbf{L},q}_{\mathbf{V}}|
\end{equation}
To upper-bound $|\mathcal{T}^{\mathbf{L},q}_{\mathbf{V}}|$ we perform counting from left to right (see Fig. \ref{packingtenycikk}). This specific counting allows a uniform handling of cases. In this simple subcase any counting would work.

\begin{equation} \label{cardialitybound}
|\mathcal{T}^{\mathbf{L},q}_{\mathbf{V}}|\le 2^{n_1 \HH_{V_1}(X)+\sum_{i=2}^{g} n_i\HH_{V_i}(\hat{X}X)+ n_{g+1} \HH_{V_{g+1}}(\hat{X}X) + n_{g+2} \HH_{V_{g+2}}(X) } \\
\end{equation}
Substituting (\ref{cardialitybound}) into (\ref{becslesalap}), the fact that $\HH_{V_i}(X)=\HH(P^{k_{i-1}})$, $3 \le i \le g$, and some algebraic rearrangement give:
\begin{align}
&\mathds{E}\left( 1^{\mathbf{L},r}_{\mathbf{V}}(\mathbf{X}_{\hat{a}}^{\hat{k}};\mathbf{X}_{a_1}^{k_1}, \dots, \mathbf{X}_{a_g}^{k_g})\right) \le subexp(n) 2^{\sum_{i=2}^{g+1} n_i (\HH_{V_i}(\hat{X}X)-\HH_{V_i}(X)-\HH_{V_i}(\hat{X})) }  2^{-l^{k_1} \HH(P^{k_1}) +n_1\HH_{V_1}(X) + n_2 \HH_{V_2}(X) }\notag\\
&\cdot 2^{-l^{k_g} \HH(P^{k_g}) +n_{g+1}\HH_{V_{g+1}}(X) + n_{g+2} \HH_{V_{g+2}}(X) } 2^{-l^{\hat{k}}\HH(P^{\hat{k}}) +  \sum_{i=2}^{g+1} n_i \HH_{V_i}(\hat{X}) }\label{becslesreszelejecase1a}\\
&=subexp(n) 2^{-\sum\limits_{i=2}^{g+1} n_i\I_{V_i}(X \wedge \hat X)} 2^{-  l^{k_1}J(V_1^{X},V_2^{X})} 2^{-l^{k_g}J(V_{g+1}^{X}, V_{g+2}^{X})} 2^{- l^{\hat{k}}J(V_2^{\hat{X}},\dots,V_{g+1}^{\hat{X}})} \label{becsleskozepecase1a}\\
&\le subexp(n) 2^{-\sum\limits_{i=2}^{g+1} n_i\I_{V_i}(X \wedge \hat X)-l^{\hat{k}}J(V_2^{\hat{X}},\dots,V_{g+1}^{\hat{X}})}.\label{becslesvegsocase1a}
\end{align}
Here in (\ref{becslesvegsocase1a}) the positivity of the Jensen-Shannon divergence is used. Inequality (\ref{becslesvegsocase1a}) implies (\ref{egysegesbecsles2}) in this subcase, under the supplementary assumption that $\mathcal{I}$ is empty. Next we prove (\ref{egysegesbecsles2}) in the general scenario of subcase 1a. Heuristically we can summarize the formal proof below that if $j \notin \mathcal{I}^*$ then in (\ref{becslesalap}) the term $l^{k_j}\HH(P^{k_j})$ is missing and in (\ref{cardialitybound}) instead of $\HH_{V_{j}}(\hat{X}X)$ the term $\HH_{V_{j}}(\hat{X}|X)$ occurs, thus the same upper-bound is obtained since
\begin{equation}
\HH_{V_i}(\hat{X}X)-\HH_{V_i}(X)-\HH_{V_i}(\hat{X})=  \HH_{V_i}(\hat{X}|X)-\HH_{V_i}(\hat{X})=-\I_{V_i}(X \wedge \hat X). \label{mutualinfo}
\end{equation}

Formally, assume first that $g \in \mathcal{I}^*$. The analogues of (\ref{becslesalap}) and (\ref{cardialitybound}) are respectively

\begin{align}
&\mathds{E}\left( 1^{\mathbf{L},q}_{\mathbf{V}}(\mathbf{X}_{\hat{a}}^{\hat{k}};\mathbf{X}_{a_1}^{k_1}, \dots, \mathbf{X}_{a_g}^{k_g})\right) \le subexp(n)\cdot 2^{-l^{\hat{k}}\HH(P^{\hat{k}})-\sum_{i \in \mathcal{I}^*}l^{k_i}\HH(P^{k_i})}\cdot |\mathcal{T}^{\mathbf{L},q}_{\mathbf{V}, \mathcal{I}}| \label{becslesalapism}\\
&|\mathcal{T}^{\mathbf{L},q}_{\mathbf{V}, \mathcal{I}}|\le 2^{n_1 \HH_{V_1}(X)+\sum_{i: i-1 \in \mathcal{I}^{*} \setminus \{g\} } n_{i} \HH_{V_i}(\hat{X}X) +\sum_{i: i-1 \in [g] \setminus \mathcal{I}^{*}} n_{i} \HH_{V_i}(\hat{X}|X)+ n_{g+1} \HH_{V_{g+1}}(\hat{X}X)+ n_{g+2} \HH_{V_{g+2}}(X) } \label{cardialitybound2ism}
\end{align}
Substituting (\ref{cardialitybound2ism}) into (\ref{becslesalapism}), (\ref{mutualinfo}) and the same algebraic rearrangement as in (\ref{becslesreszelejecase1a}) give identical upper-bound to the one in (\ref{becsleskozepecase1a}).

In case of $g \notin \mathcal{I}^*$ the analogues of (\ref{becslesalap}) and (\ref{cardialitybound}) are respectively
\begin{align}
&\mathds{E}\left( 1^{\mathbf{L},q}_{\mathbf{V}}(\mathbf{X}_{\hat{a}}^{\hat{k}};\mathbf{X}_{a_1}^{k_1}, \dots, \mathbf{X}_{a_g}^{k_g})\right) \le subexp(n)\cdot 2^{-l^{\hat{k}}\HH(P^{\hat{k}})-\sum_{i \in \mathcal{I}^*}l^{k_i}\HH(P^{k_i})}\cdot |\mathcal{T}^{\mathbf{L},q}_{\mathbf{V}, \mathcal{I}}| \label{becslesalapismm}\\
&|\mathcal{T}^{\mathbf{L},q}_{\mathbf{V}, \mathcal{I}}|\le 2^{n_1 \HH_{V_1}(X)+\sum_{i: i-1 \in \mathcal{I}^{*}} n_{i} \HH_{V_i}(\hat{X}X) +\sum_{i: i-1 \in [g-1] \setminus \mathcal{I}^{*}} n_{i} \HH_{V_i}(\hat{X}|X)+ n_{g+1} \HH_{V_{g+1}}(\hat{X}|X) } \label{cardialitybound2ismm}
\end{align}
Now substituting (\ref{cardialitybound2ismm}) into (\ref{becslesalapismm}) and proceeding similarly as before give upper-bound (\ref{becsleskozepecase1a}) without $2^{-l^{k_g}J(V_{g+1}^{X}, V_{g+2}^{X})}$ which is omitted in the next step.

This completes the proof of (\ref{egysegesbecsles2}) in subcase 1a. Note that the argument which allowed the proof without the supplementary assumption works also in other subcases. Hence, from now on we assume that no equality  $x_i=x_j$, $i,j \in [g]$, is in $\mathcal{I}$ except the equality $x_1=x_g$ in subcase 1d.

\emph{SUBCASE 1b}: According to the last paragraph of subcase 1a, it can be assumed that $\mathcal{I}=\{\hat{\mathbf{x}}=\mathbf{x}_1\}$. We can assume also that there exists a collection of sequences $(\hat{\mathbf{x}},\mathbf{x}_{1}, \dots \mathbf{x}_{g})$ with $\hat{\mathbf{x}} \in \mathcal{T}_{P^{\hat{k}}}^{l^{\hat{k}}}(\gamma_n)$ and $\mathbf{x}_{i} \in \mathcal{T}_{P^{k_i}}^{l^{k_i}}(\gamma_n)$ for all $i \in [g]$ with $\mathds 1_{\mathbf{V}, \mathcal{I}}^{\mathbf{L},q} (\hat{\mathbf{x}};\mathbf{x}_{1},\dots, \mathbf{x}_{g})=1$,  otherwise $\mathds{E}\left( 1^{\mathbf{L},q}_{\mathbf{V}}(\mathbf{X}_{\hat{a}}^{\hat{k}};\mathbf{X}_{a_1}^{k_1}, \dots, \mathbf{X}_{a_g}^{k_g})\right)$ is equal to $0$. This assumption implies that
\begin{equation} \label{expurgalaseredmenye}
2^{n_2 \I_{V_2}(X \wedge \hat{X})} < subexp(n)
\end{equation}
because if $n_2 < (\log n)^2$ then (\ref{expurgalaseredmenye}) is immediate, while otherwise $\I_{V_2}(X \wedge \hat{X})< \gamma_n$ which also implies (\ref{expurgalaseredmenye}). Then using (\ref{basicfact2}) and (\ref{expurgsize}) we get that
\begin{equation} \label{becslesalap1b}
\mathds{E}\left( 1^{\mathbf{L},q}_{\mathbf{V}}(\mathbf{X}_{\hat{a}}^{\hat{k}};\mathbf{X}_{a_1}^{k_1}, \dots, \mathbf{X}_{a_g}^{k_g})\right) \le subexp(n)\cdot 2^{-l^{\hat{k}}\HH(P^{\hat{k}})-\sum_{i=2}^{g} l^{k_i} \HH(P^{k_i})}\cdot |\mathcal{T}^{\mathbf{L},q}_{\mathbf{V}, \mathcal{I}}|
\end{equation}
To upper-bound $|\mathcal{T}^{\mathbf{L},q}_{\mathbf{V}, \mathcal{I}}|$ we perform counting again from left to right (see Fig. \ref{packingtenycikk}) but we skip the first block, and write $n_2 \HH_{V_2}(\hat{X})$ instead of $n_2 \HH_{V_2}(\hat{X}X)$ related to the second block.
\begin{equation} \label{cardialitybound1b}
|\mathcal{T}^{\mathbf{L},q}_{\mathbf{V}, \mathcal{I}}|\le 2^{n_2 \HH_{V_2}(\hat{X})+\sum_{i=3}^{g} n_i\HH_{V_i}(\hat{X}X)+ n_{g+1} \HH_{V_{g+1}}(\hat{X}X) + n_{g+2} \HH_{V_{g+2}}(X) } \\
\end{equation}
Substituting (\ref{cardialitybound1b}) into (\ref{becslesalap1b}), the fact that $\HH_{V_i}(X)=\HH(P^{k_{i-1}})$, $3 \le i \le g$, and performing the same algebraic rearrangement as before give:
\begin{align}
&\mathds{E}\left( 1^{\mathbf{L},q}_{\mathbf{V}}(\mathbf{X}_{\hat{a}}^{\hat{k}};\mathbf{X}_{a_1}^{k_1}, \dots, \mathbf{X}_{a_g}^{k_g})\right) \le subexp(n) 2^{-\sum\limits_{i=3}^{g+1} n_i\I_{V_i}(X \wedge \hat X)} 2^{-l^{k_g}J(V_{g+1}^{X}, V_{g+2}^{X})} 2^{- l^{\hat{k}}J(V_2^{\hat{X}},\dots,V_{g+1}^{\hat{X}})} \label{becsleskozepecase1b}\\
&\le subexp(n) 2^{-\sum\limits_{i=2}^{g+1} n_i\I_{V_i}(X \wedge \hat X)-l^{\hat{k}}J(V_2^{\hat{X}},\dots,V_{g+1}^{\hat{X}})}.\label{becslesvegsocase1b}
\end{align}
Here in (\ref{becslesvegsocase1b}) the positivity of the Jensen-Shannon divergence and (\ref{expurgalaseredmenye}) are used. Inequality (\ref{becslesvegsocase1b}) implies (\ref{egysegesbecsles2}).

\emph{SUBCASE 1c}: According to the last paragraph of subcase 1a it can be assumed that $\mathcal{I}=\{\hat{\mathbf{x}}=\mathbf{x}_g\}$. Now using again (\ref{basicfact2}) and (\ref{expurgsize}) we get that
\begin{equation} \label{becslesalap1c}
\mathds{E}\left( 1^{\mathbf{L},q}_{\mathbf{V}}(\mathbf{X}_{\hat{a}}^{\hat{k}};\mathbf{X}_{a_1}^{k_1}, \dots, \mathbf{X}_{a_g}^{k_g})\right) \le subexp(n)\cdot 2^{-l^{\hat{k}}\HH(P^{\hat{k}})-\sum_{i=1}^{g-1}l^{k_i}\HH(P^{k_i})}\cdot |\mathcal{T}^{\mathbf{L},q}_{\mathbf{V}, \mathcal{I}}|
\end{equation}

\begin{figure}[h]
\begin{center}
\includegraphics[scale=1.5]{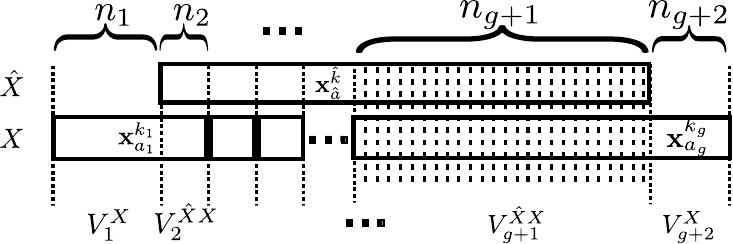}
\caption{Further division of the $(g+1)$'th subblock in subcases (1c) and (1d)}
\label{packingtenycikkmodtovabb}
\end{center}
\end{figure}

To upper-bound $|\mathcal{T}^{\mathbf{L},q}_{\mathbf{V}, \mathcal{I}}|$ we divide the subblock of index $g+1$ into consecutive subblocks of length $n_{g+1,i}=(\log n)^2$ except perhaps for the last subblock that has length $n_{g+1,s}\le (\log n)^2$ (see Fig. \ref{packingtenycikkmodtovabb}). Again we perform the counting from left to right.
\begin{align} \label{cardialitybound1c}
&|\mathcal{T}^{\mathbf{L},q}_{\mathbf{V},\mathcal{I}}|\le \sum_{\sumfrac{V_{g+1,i}\in \mathcal{P}^{n_i}(\mathcal{X}\times \mathcal{X}), i \in [s]}{V_{g+1,1} \oplus \dots \oplus V_{g+1,s}=V_{g+1}}} 2^{n_1 \HH_{V_1}(X)+\sum_{i=2}^{g} n_i\HH_{V_i}(\hat{X}X)+\sum_{i=1}^{s} n_{g+1,i} \HH_{V_{g+1,i}}(\hat{X}|X)  } \\
&\le subexp(n) 2^{n_1 \HH_{V_1}(X)+\sum_{i=2}^{g} n_{i} \HH_{V_i}(\hat{X}X) + n_{g+1} \HH_{V_{g+1}}(\hat{X}|X)  } \label{cardialitybound21c}
\end{align}
Here in (\ref{cardialitybound1c}) the sum is over subtype sequences corresponding to the division in Fig. \ref{packingtenycikkmodtovabb}, where $V_{g+1,1}\oplus \dots \oplus V_{g+1,s}=V_{g+1}$. In (\ref{cardialitybound21c}) we used the concavity of entropy and the fact that the number of subtype sequences in the sum is subexponential. Substituting (\ref{cardialitybound21c}) into (\ref{becslesalap1c}), the fact that $\HH_{V_i}(X)=\HH(P^{k_{i-1}})$, $3 \le i \le g$, and the same algebraic rearrangement as before give:
\begin{align}
&\mathds{E}\left( 1^{\mathbf{L},q}_{\mathbf{V}}(\mathbf{X}_{\hat{a}}^{\hat{k}};\mathbf{X}_{a_1}^{k_1}, \dots, \mathbf{X}_{a_g}^{k_g})\right) \le subexp(n) 2^{\sum_{i=2}^{g+1} n_i (\HH_{V_i}(\hat{X}X)-\HH_{V_i}(X)-\HH_{V_i}(\hat{X})) }  2^{-l^{k_1} \HH(P^{k_1}) +n_1\HH_{V_1}(X) + n_2 \HH_{V_2}(X) }\notag\\
&\cdot 2^{-l^{\hat{k}}\HH(P^{\hat{k}}) +  \sum_{i=2}^{g+1} n_i \HH_{V_i}(\hat{X}) }\label{becslesreszelejecase1c}\\
&=subexp(n) 2^{-\sum\limits_{i=2}^{g+1} n_i\I_{V_i}(X \wedge \hat X)} 2^{-  l^{k_1}J(V_1^{X},V_2^{X})} 2^{- l^{\hat{k}}J(V_2^{\hat{X}},\dots,V_{g+1}^{\hat{X}})} \label{becsleskozepecase1c}\\
&\le subexp(n) 2^{-\sum\limits_{i=2}^{g+1} n_i\I_{V_i}(X \wedge \hat X)-l^{\hat{k}}J(V_2^{\hat{X}},\dots,V_{g+1}^{\hat{X}})}.\label{becslesvegsocase1c}
\end{align}
Here in (\ref{becslesvegsocase1c}) the positivity of the Jensen-Shannon divergence is used. Inequality (\ref{becslesvegsocase1c}) implies (\ref{egysegesbecsles2}).

\emph{SUBCASE 1d}: In this subcase we have to combine the methods of previous subcases. According to the last paragraph of subcase 1a it can be assumed that $\mathcal{I}=\{\hat{\mathbf{x}}=\mathbf{x}_1,\hat{\mathbf{x}}=\mathbf{x}_g, \mathbf{x}_1=\mathbf{x}_g\}$. We can assume also that there exists a collection of sequences $(\hat{\mathbf{x}},\mathbf{x}_{1}, \dots \mathbf{x}_{g})$ with $\hat{\mathbf{x}} \in \mathcal{T}_{P^{\hat{k}}}^{l^{\hat{k}}}(\gamma_n)$ and $\mathbf{x}_{i} \in \mathcal{T}_{P^{k_i}}^{l^{k_i}}(\gamma_n)$ for all $i \in [g]$ with $\mathds 1_{\mathbf{V}, \mathcal{I}}^{\mathbf{L},q} (\hat{\mathbf{x}};\mathbf{x}_{1},\dots, \mathbf{x}_{g})=1$,  otherwise $\mathds{E}\left( 1^{\mathbf{L},q}_{\mathbf{V}}(\mathbf{X}_{\hat{a}}^{\hat{k}};\mathbf{X}_{a_1}^{k_1}, \dots, \mathbf{X}_{a_g}^{k_g})\right)$ is equal to $0$. This assumption implies as in subcase 1b that
\begin{equation} \label{expurgalaseredmenye1d}
2^{n_2 \I_{V_2}(X \wedge \hat{X})} < subexp(n).
\end{equation}
Now using again (\ref{basicfact2}) we get that
\begin{equation} \label{becslesalap1d}
\mathds{E}\left( 1^{\mathbf{L},q}_{\mathbf{V}}(\mathbf{X}_{\hat{a}}^{\hat{k}};\mathbf{X}_{a_1}^{k_1}, \dots, \mathbf{X}_{a_g}^{k_g})\right) \le subexp(n)\cdot 2^{-l^{\hat{k}}\HH(P^{\hat{k}})-\sum_{i=2}^{g-1}l^{k_i}\HH(P^{k_i})}\cdot |\mathcal{T}^{\mathbf{L},q}_{\mathbf{V}, \mathcal{I}}|
\end{equation}
To upper-bound $|\mathcal{T}^{\mathbf{L},q}_{\mathbf{V}, \mathcal{I}}|$ we divide the subblock corresponding to $V_{g+1}$ into consecutive subblocks of length $(\log n)^2$ as in subcase 1c (see Fig. \ref{packingtenycikkmodtovabb}) and as in subcase (1b) we perform the counting from left to right but we skip the first block and write $n_2 \HH_{V_2}(\hat{X})$ instead of $n_2 \HH_{V_2}(\hat{X}X)$ related to the second block. Using the same arguments as in subcases 1b and 1c we get
\begin{align} \label{cardialitybound1d}
&|\mathcal{T}^{\mathbf{L},q}_{\mathbf{V},\mathcal{I}}|\le \sum_{\sumfrac{V_{g+1,i}\in \mathcal{P}^{n_i}(\mathcal{X}\times \mathcal{X}), i \in [s]}{V_{g+1,1} \oplus \dots \oplus V_{g+1,s}=V_{g+1}}} 2^{n_2 \HH_{V_2}(\hat{X})+\sum_{i=3}^{g} n_i\HH_{V_i}(\hat{X}X)+\sum_{i=1}^{s} n_{g+1,i} \HH_{V_{g+1,i}}(\hat{X}|X)  } \\
&\le subexp(n) 2^{n_2 \HH_{V_2}(\hat{X})+\sum_{i=3}^{g} n_{i} \HH_{V_i}(\hat{X}X) + n_{g+1} \HH_{V_{g+1}}(\hat{X}|X)  } \label{cardialitybound21d} \\
&\mathds{E}\left( 1^{\mathbf{L},q}_{\mathbf{V}}(\mathbf{X}_{\hat{a}}^{\hat{k}};\mathbf{X}_{a_1}^{k_1}, \dots, \mathbf{X}_{a_g}^{k_g})\right) \le subexp(n) 2^{-\sum\limits_{i=3}^{g+1} n_i\I_{V_i}(X \wedge \hat X)} 2^{- l^{\hat{k}}J(V_2^{\hat{X}},\dots,V_{g+1}^{\hat{X}})} \label{becsleskozepecase1d}\\
&\le subexp(n) 2^{-\sum\limits_{i=2}^{g+1} n_i\I_{V_i}(X \wedge \hat X)-l^{\hat{k}}J(V_2^{\hat{X}},\dots,V_{g+1}^{\hat{X}})}.\label{becslesvegsocase1d}
\end{align}
Inequality (\ref{becslesvegsocase1d}) implies (\ref{egysegesbecsles2}).

\begin{figure}[h]
\begin{center}
\includegraphics[scale=1.5]{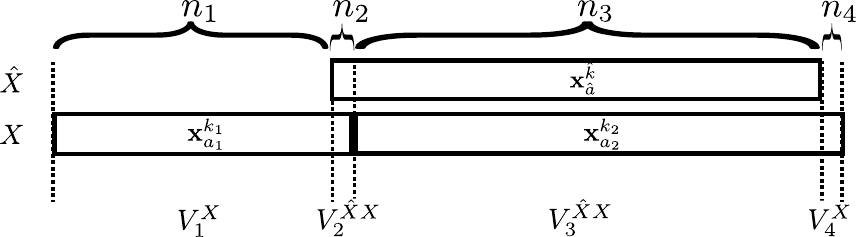}
\caption{Illustration for case (2)}
\label{kepmarkov1}
\end{center}
\end{figure}

\emph{CASE 2}: In this case $g = 2$ holds if $n$ is large enough. Moreover, by symmetry it can be assumed that $n_2 \le (\log n)^2$ and $n_3 \ge n-(\log n)^2$ (See Fig. \ref{kepmarkov1}). Here also we separately investigate the same four subcases.

\emph{SUBCASES 2a and 2b}: The proofs are identical to the proofs of subcases 1a and 1b respectively.

\emph{SUBCASE 2c}: Here $\mathcal{I}=\{\hat{\mathbf{x}}=\mathbf{x}_1\}$. Using again (\ref{basicfact2}) we get that
\begin{equation} \label{becslesalap2c}
\mathds{E}\left( 1^{\mathbf{L},q}_{\mathbf{V}}(\mathbf{X}_{\hat{a}}^{\hat{k}};\mathbf{X}_{a_1}^{k_1}, \dots, \mathbf{X}_{a_g}^{k_g})\right) \le subexp(n)\cdot 2^{-l^{\hat{k}}\HH(P^{\hat{k}})-l^{k_1}\HH(P^{k_1})}\cdot |\mathcal{T}^{\mathbf{L},q}_{\mathbf{V}, \mathcal{I}}|
\end{equation}
\begin{figure}[h]
\begin{center}
\includegraphics[scale=1.5]{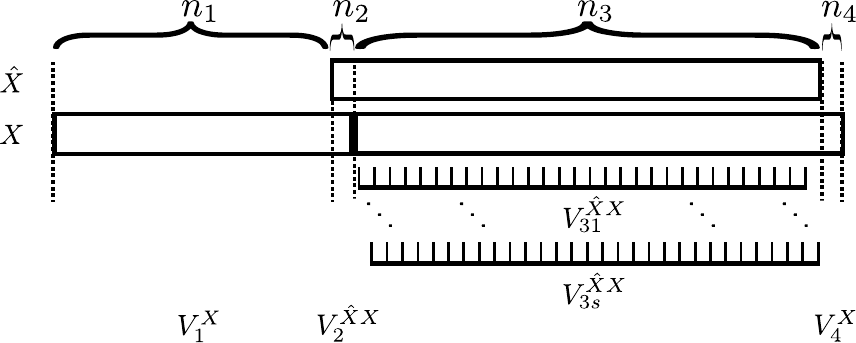}
\caption{Further division of the third subblock in subcases 2c and 2d}
\label{kepmarkov2}
\end{center}
\end{figure}
To upper-bound $|\mathcal{T}^{\mathbf{L},q}_{\mathbf{V}, \mathcal{I}}|$ we divide the third block into $n_2$ "virtual subblocks" consisting of non-consecutive elements: the first virtual subblock corresponds to indices $(1,1+n_2,\dots)$, the second one corresponds to indices $(2,2+n_2,\dots)$, $\dots$, the last corresponds to indices $(n_2,2n_2,\dots)$ (see Fig. \ref{kepmarkov2}). Let $n_{3,1}$, \dots, $n_{3,s}$ denote the lengths of these "subblocks" (note that $|n_{3,i}-n_{3,j}|\le 1$ for all $i,j \in [s]$). Then:
\begin{align} \label{cardialitybound2c}
&|\mathcal{T}^{\mathbf{L},q}_{\mathbf{V},\mathcal{I}}|\le \sum_{\sumfrac{V_{3,i}\in \mathcal{P}^{n_{3,i}}(\mathcal{X}\times \mathcal{X}), i \in [s]}{V_{3,1} \oplus \dots \oplus V_{3,s}=V_{3}}} 2^{n_1 \HH_{V_1}(X)+ n_2\HH_{V_2}(\hat{X}X)+\sum_{i=1}^{s} n_{3,i} \HH_{V_{3,i}}(\hat{X}|X)  } \\
&\le subexp(n) 2^{n_1 \HH_{V_1}(X)+ n_{2} \HH_{V_2}(\hat{X}X) + n_{3} \HH_{V_{3}}(\hat{X}|X)  } \label{cardialitybound22c}
\end{align}
Here in (\ref{cardialitybound2c}) inequality (\ref{markovrabecsles}) is used and the sum is over type sequences corresponding to the division in Fig. \ref{kepmarkov2}. These $V_{3,1},\dots, V_{3,s}$ have convex combination $V_{3}$. In (\ref{cardialitybound22c}) again the concavity of the entropy and the fact that the number of subtype sequences in the sum is subexponential in $n$ are used. Substituting (\ref{cardialitybound22c}) into (\ref{becslesalap2c}) and the same algebraic rearrangement as before give:
\begin{align}
&\mathds{E}\left( 1^{\mathbf{L},q}_{\mathbf{V}}(\mathbf{X}_{\hat{a}}^{\hat{k}};\mathbf{X}_{a_1}^{k_1}, \dots, \mathbf{X}_{a_g}^{k_g})\right) \le subexp(n) 2^{-\sum\limits_{i=2}^{3} n_i\I_{V_i}(X \wedge \hat X)} 2^{-  l^{k_1}J(V_1^{X},V_2^{X})} 2^{- l^{\hat{k}}J(V_2^{\hat{X}},V_{3}^{\hat{X}})} \label{becsleskozepecase2c}\\
&\le subexp(n) 2^{-\sum\limits_{i=2}^{3} n_i\I_{V_i}(X \wedge \hat X)-l^{\hat{k}}J(V_2^{\hat{X}},V_{3}^{\hat{X}})}.\label{becslesvegsocase2c}
\end{align}
Here in (\ref{becslesvegsocase2c}) the positivity of the Jensen-Shannon divergence is used again. Inequality (\ref{becslesvegsocase2c}) implies (\ref{egysegesbecsles2}).

\emph{SUBCASE 2d}: Here $\mathcal{I}=\{\hat{\mathbf{x}}=\mathbf{x}_1, \hat{\mathbf{x}}=\mathbf{x}_2,\mathbf{x}_1=\mathbf{x}_2 \}$. Note that (\ref{expurgalaseredmenye1d}) trivially holds in this case. Using again (\ref{basicfact2}) we get that
\begin{equation} \label{becslesalap2d}
\mathds{E}\left( 1^{\mathbf{L},q}_{\mathbf{V}}(\mathbf{X}_{\hat{a}}^{\hat{k}};\mathbf{X}_{a_1}^{k_1}, \dots, \mathbf{X}_{a_g}^{k_g})\right) \le subexp(n)\cdot 2^{-l^{\hat{k}}\HH(P^{\hat{k}})}\cdot |\mathcal{T}^{\mathbf{L},q}_{\mathbf{V}, \mathcal{I}}|
\end{equation}
To upper-bound $|\mathcal{T}^{\mathbf{L},q}_{\mathbf{V}, \mathcal{I}}|$ we divide the third block into $n_2$ "subblocks" consisting of non-consecutive elements as in subcase (2c) (see Fig. \ref{kepmarkov2}) and we skip the first block and write $n_2\HH_{V_2}(\hat{X})$ instead of $n_2\HH_{V_2}(\hat{X}X)$ related to the second block. Using the same argument as in subcases 1d and 2c we get:
\begin{align} \label{cardialitybound2d}
&|\mathcal{T}^{\mathbf{L},q}_{\mathbf{V},\mathcal{I}}|\le \sum_{\sumfrac{V_{3,i}\in \mathcal{P}^{n_{3,i}}(\mathcal{X}\times \mathcal{X}), i \in [s]}{V_{3,1} \oplus \dots \oplus V_{3,s}=V_{3}}} 2^{n_2\HH_{V_2}(\hat{X})+\sum_{i=1}^{s} n_{3,i} \HH_{V_{3,i}}(\hat{X}|X)  } \\
&\le subexp(n) 2^{n_{2} \HH_{V_2}(\hat{X}) + n_{3} \HH_{V_{3}}(\hat{X}|X)  } \label{cardialitybound22d} \\
&\mathds{E}\left( 1^{\mathbf{L},q}_{\mathbf{V}}(\mathbf{X}_{\hat{a}}^{\hat{k}};\mathbf{X}_{a_1}^{k_1}, \dots, \mathbf{X}_{a_g}^{k_g})\right) \le subexp(n) 2^{- n_3\I_{V_3}(X \wedge \hat X)} 2^{- l^{\hat{k}}J(V_2^{\hat{X}},V_{3}^{\hat{X}})} \label{becsleskozepecase2d}\\
&\le subexp(n) 2^{-\sum\limits_{i=2}^{3} n_i\I_{V_i}(X \wedge \hat X)-l^{\hat{k}}J(V_2^{\hat{X}},V_{3}^{\hat{X}})}.\label{becslesvegsocase2d}
\end{align}
Inequality (\ref{becslesvegsocase2d}) implies (\ref{egysegesbecsles2}).

\begin{Rem}
In the proof above the different divisions shown on Fig. \ref{packingtenycikkmodtovabb} and Fig. \ref{kepmarkov2} ensure that the numbers of terms in the corresponding sums are subexponential.
\end{Rem}

\section*{Acknowledgment}
We would like to thank Prof. Imre Csiszár for his help and advice. We also thank the support of the Hungarian National Research Development and Innovation Office Grant K105840 and the MTA-BME Stochastics Research Group.

\end{document}